\documentclass[a4paper,12pt]{article}
\usepackage{graphicx}
\usepackage{amssymb}
\usepackage{epstopdf}
\usepackage{bm}
\usepackage[a4paper, total={6.5in, 9in}]{geometry}
\usepackage{hyperref}
\usepackage{xcolor}
\usepackage{dsfont}
\usepackage{eurosym}
\usepackage{hyperref}
\begin{document}
\title{\LARGE How Covid mobility restrictions modified the population of investors in Italian stock markets\footnote{Disclaimer: This paper represents the personal opinions of the authors and do not bind the membership organization in any way.}}
\author{Paola Deriu\thanks{Consob, Italy.\iffalse The opinions expressed are personal and do not bind the membership organization in any way.\fi}, Fabrizio Lillo\thanks{Scuola Normale Superiore, Pisa and Dipartimento di Matematica, Universit\`a di Bologna, Italy}, Piero Mazzarisi\thanks{Scuola Normale Superiore, Pisa, Italy},\\Francesca Medda\thanks{Consob, Italy.\iffalse The opinions expressed are personal and do not bind the membership organization in any way.\fi}, Adele Ravagnani\thanks{Scuola Normale Superiore, Pisa, Italy}, Antonio Russo\thanks{Consob, Italy.\iffalse The opinions expressed are personal and do not bind the membership organization in any way.\fi}}

\maketitle

\abstract{This paper investigates how Covid mobility restrictions impacted the population of investors of the Italian stock market. The analysis tracks the trading activity of individual investors in  Italian stocks in the period January 2019-September 2021, investigating how their composition and the trading activity changed around the Covid-19 lockdown period (March 9 - May 19, 2020) and more generally in the period of the pandemic. The results pinpoint that the lockdown restriction was accompanied by a surge in interest toward stock market, as testified by the trading volume by households. Given the generically falling prices during the lockdown, the households, which are typically contrarian, were net buyers, even if less than expected from their trading activity in 2019. This can be explained by the arrival, during the lockdown, of a group of $\sim 185$k new investors (i.e. which had never traded since January 2019) which were on average ten year younger and with a larger fraction of males than the pre-lockdown investors. By looking at the gross P\&L, there is clear evidence that these new investors were more skilled in trading. There are thus 
indications that the lockdown, and more generally the Covid pandemic, created a sort of regime change in the population of financial investors.
\vspace{0.1cm}}

\section{Introduction}

Undoubtedly, Covid pandemic represented one of the major shocks the entire planet has experienced in the last decades. Our lives have been severely affected in the way we work, travel, interact with other persons, essentially in all aspects of our existence. Some of these changes were mainly focused in the periods when mobility and interaction restrictions were in place to avoid the spreading of the contagion, while others had a permanent effect. We are still not out of the global pandemic phase, but we can easily foresee that, even when all the restrictions will be removed, our lives will be very different from what they were in January 2020. The pandemic has accelerated some transitions that were 
predicted to happen in longer time spans. Especially the activity related to the digital transition, remote working, social interaction, and many others experienced a fast evolution toward a much wider adoption.

Very likely, finance and access to financial markets have been strongly modified during the Covid period, accelerating a trend already in place and related to the digital transition in Finance (Fintech, access to trading platforms, cryptocurrencies, etc).
However, a proper quantification of the effects of Covid restrictions on the behaviour of retail investors is still missing. As discussed below, some very recent studies have started to investigate this subject, but, in our opinion they lack some important dimensions that instead are properly taken into account in this paper. Precisely, this study possesses the following characteristics: (1) access to data describing the behaviour of {\it all} the {\it actual}\footnote{As we will review below, some recent literature uses portion of data coming from specific providers (e.g. Robinhood in the US) or infer retail trades from public market data by analyzing the volume traded outside registered exchanges.} investors in all Italian stocks; (2) wide investigated time span (January 2019 - September 2021), which allows to disentangle {\it transient} from {\it permanent} effects on the behaviour and composition of household investors of the main mobility restrictions in March-May 2020; (3) use of {\it dynamical} machine learning methods to identify the modification of household investors' strategy to a rapidly changing financial environment, as the high volatility phase experienced by many assets during the lockdown period.

Our main findings can be summarized as follows: during (and sometimes slightly before) the first lockdown, Italian stocks experienced a very significant increase in the traded volume by households. Prices were typically falling and households, being typically contrarian - i.e. investors going against prevailing market trends, by selling when prices increases, and buying when prices are decreasing - were net buyers; however the dynamical kernel regression model clarifies that households were significantly less contrarian than one would have expected from their trading activity before the lockdown. This can be explained by one of the most important findings of this study, namely the entrance, during the lockdown, of a group of $\sim 185k$ new investors (i.e. which had never traded since January 2019). By looking at their demographic characteristics, we find that these newcomers are on average ten year younger and with a larger fraction of males than the
pre-lockdown investors. Moreover, the analysis of their gross P\&L indicates that these new investors were skilled in trading. Finally, by comparing the market investors composition during the relatively long investigated time span, strong indications are found that the lockdown, and more generally the Covid period, has permanently modified the population of investors, as testified by the fact that the vast majority of newcomers were still active one year and half after the lockdown and by the increased fraction of volume traded by households (with respect to the one traded by institutional investors). 
Thus, data indicate a significant change in the composition and in the trading activity of retail investors.

{\bf Literature review.} Few weeks after the pandemic outbreak, the French regulator AMF \cite{amf} issued a short report where, analysing retail investors, they found that purchases of French equities by retail clients increased fourfold in March 2020, and overall volumes tripled. Among the many equity buyers during this period, a significant proportion were new clients and these new clients are between 10 and 15 years younger than regular investors in French equities. The AMF considered only the period February 24, 2020 to April 3, 2020 and found that 90\% of the long positions were still held at the end of the six crisis weeks analysed, with only 10\% having been subject to opportunistic trading or asset reallocation. Our paper has some similarities with the AMF study, the main differences being the much longer investigated time span, allowing to identify permanent regime shifts in the market composition and a better assessment of the trading skills of the newcomers, and the dynamic analysis of the relation between prices and retail trading. 
 
An extensive analysis about the increase of trading activity due to pandemic has been presented by Chiah and Zhong \cite{chiah}, which pointed out a large spike in trading volume in 37 international equity markets in the period of lockdown and found that investors have traded more heavily in wealthier nations, in particular the ones with stronger protection of legal rights, better governance systems, and greater gambling opportunities. Ozik et al. \cite{ozik} studied the impact of retail investors on stock liquidity during the Covid-19 lockdown in Spring 2020 and showed a sharp increase in retail trading, especially among stocks with high COVID-19-related media coverage and conclude that ``access to financial markets facilitated by fintech innovations to trading platforms, along with ample free time, are significant determinants of retail investor stock-market participation." Note that their analysis is limited to a subset of investors as monitored by looking at ``hourly snapshots of Robinhood popularity metrics which represent the number of unique Robinhood user accounts holding at least one share of the stock."  The investigated period is January-October 2020; thus, they were not able (as the authors of the AMF report) to track the persistence of new investors in the market far from the lockdown. The focus of their paper is on the effect of retail investor volume on the increased liquidity in the market.

Many European regulators have also highlighted the increased interest of retail investors on financial markets since the outburst of the pandemic. For example, the Italian regulator (CONSOB) has published analyses on queries to Google from Italy of terms related to stock markets and online trading and on trading activity of Italian investors from 2019 onwards based on supervisory data (see, e.g., the latest 2021 CONSOB report \cite{consobreport2021}). The analyses show that interest in trading activity of Italian investors has risen substantially over the time span covered. Trades recorded a spike in March 2020 both in Italian equities and (although to a lesser extent) bonds and mutual funds, when positive net purchases prevailed, marking a partial reversal in the following months. Male investors and middle-aged investors has continued to be the most active over time, although the share of younger investors has been rising too. More evidence also from other European regulators can be found in ESMA (2002) \cite{esma}, focusing on the development of retail risk indicators for the EU single market.

Several recent studies \cite{Odean2021} perform analyses of the profitability and trading behaviour of retail investors by inferring from public data (e.g. TAQ database) the trades which are likely involving households, being executed outside regulated exchanges, as proposed by \cite{Boehmer}. In contrast our study is based on transaction reporting data (see Section \ref{sec:data}) where the trades by retail and institutional investors are unambiguously identified and moreover the data contains the whole universe of trades of Italian stocks.  

A database similar to the one used in this paper (even if collected for different reason and investigated in periods much before the Covid) is the one of the Finnish stock market, which has been investigated in several papers. For example, \cite{overconfidence} finds evidence of overconfidence and sensation seeking in the trading behaviour of young households.  We will discuss later if some of our findings on the arrival of new investors can be interpreted in light of these behavioral characteristics of households. \cite{news} finds that for households  trading decisions volatility and returns are much more relevant than the number of news articles and news sentiment, respectively. Clustering of agents has been proposed in \cite{svn} and applied to the study of extraordinary events, as IPO, in \cite{baltakiene}.

 The literature on the determinants of household trading motivations and performance is vast and cannot be fully reviewed here. There is strong evidence that in aggregate retail investors earn poor returns at short horizons, e.g. less than a month (as well as longer horizons) \cite{odean,Barber}. A series of behavioural phenomena have been proposed for this fact, including overconfidence and sensation seeking \cite{overconfidence,overconfidence2}, searching for entertainment \cite{dorn}, urge to gamble \cite{chiah2,gao}, and hope to gain social status \cite{han}. Despite the purpose of this paper is not to discriminate among these alternatives, it is evident that the exceptional conditions in which all the households were forces to stay, as a consequence of the mobility restrictions due to Covid, might have exacerbated some of these drivers (for example gambling), contributing to explain the surge in retail trading volume and the arrival of many new investors. 
 
 Our analysis can be seen as a contribution to the wider topic of the increasing importance of household in financial markets, as discussed, for example, in a recent article of the Financial Times \cite{ft}.

\medskip

{\bf Paper organization.}  The paper is organized as follows. Section \ref{sec:covid} briefly outlines the main events in Italy due to the Covid pandemic and the main measures taken by the Italian government to limit the mobility in the attempt to contain the contagion, while section \ref{sec:data} presents the dataset used in this study. Section \ref{sec:modify} presents the results quantifying how the investors modified their trading activity during and after the lockdown. Section \ref{sec:new} investigates the population of new investors which entered the market during the first lockdown, characterizing their demographic characteristics (age and gender) as well as investigating the profitability of their trading. Section \ref{sec:shift} studies whether the lockdown and more generally the Covid pandemic has modified the composition of the population of traders of Italian stocks. Finally, in Section \ref{sec:conclusion} some conclusions are drawn as well as suggestions for some further research.

\section{Covid-19 in Italy and its management}\label{sec:covid}

As it is well known, Italy was the first western country to be severely hit by the Covid-19 pandemic and to put in place localized and nationwide restrictions. The first cases were documented on January 30, 2020, while between February 23  and March 8 the first partial lockdowns were imposed in some municipalities of Northern Italy. Then, the first nationwide lockdown was imposed from March 9 to May 18 (and this period will be investigated in detail in this paper) even if already on May 4 some very partial reopening were allowed. While during the summer 2020 the mobility restrictions were significantly relaxed, on October 13, 2020 several administrative orders reintroduced a severe anti-contagion policy, resulting in local lockdowns at a regional level depending on some epidemic indicators. Moreover starting from November 3, 2020, the twenty Italian regions were dynamically grouped into four categories depending on the severity of contagion and consequently different restrictions are imposed. The peak of restrictions in this period was recorded between December 24, 2020 and mid-February 2021, creating a sort of second lockdown, even if less homogeneous than the first one.

\begin{figure}
\begin{center}
        \includegraphics[width=0.45\textwidth]{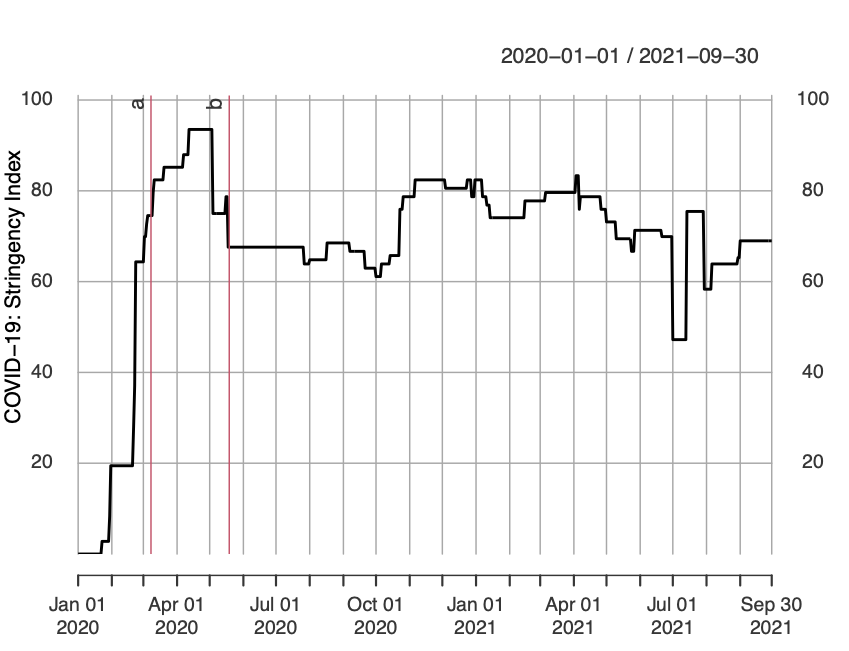}
        \includegraphics[width=0.45\textwidth]{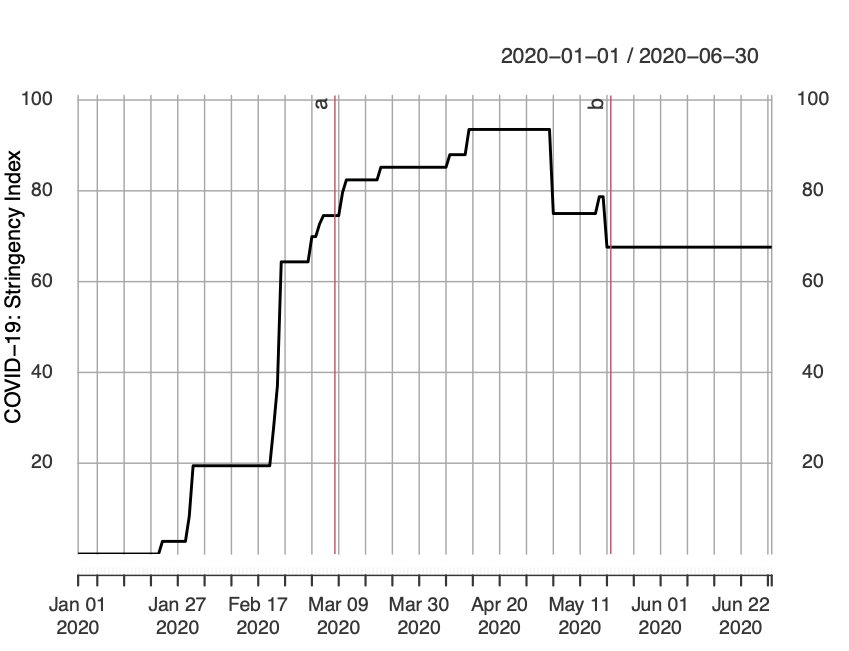}
	\caption{Left. Stringency index \cite{hale}, measuring the strenght of mobility restrictions, for Italy in the period  January 1, 2020 to September 30, 2021. Right Zoom on the first semester of 2020. The red vertical lines indicate the beginning and end of the first lockdown. The index is a composite measure based on nine response indicators rescaled to a value from $0$ to $100$ ($100$ = strictest).}\label{fig:stringency}
\end{center}
\end{figure}

In order to quantify the level of restrictions imposed by the Italian government in response to the Covid pandemic, we make use of the {\it stringency index} recently introduced \cite{hale} at Oxford University. According to their definition ``the stringency index is a composite measure based on nine response indicators including school closures, workplace closures, and travel bans, rescaled to a value from $0$ to $100$ ($100$ = strictest)". The index for more than $180$ countries is available at \href{https://www.bsg.ox.ac.uk/research/research-projects/covid-19-government-response-tracker/}{https://www.bsg.ox.ac.uk/research/research-projects/covid-19-government-response-tracker} and we extracted it for Italy.

Figure \ref{fig:stringency} shows the dynamics of the stringency index in Italy for the period January 1, 2020 (first available date) to September 30, 2021. The zoom on the first semester of 2020 (right panel) indicate that several significant restrictions were in place in the weeks before the official lockdown (vertical red lines). As we will see below this might explain why the increased interest of household toward financial markets started before the official lockdown.

\section{Dataset}\label{sec:data}

\begin{figure}
\begin{center}
        \includegraphics[width=0.45\textwidth,angle=0]{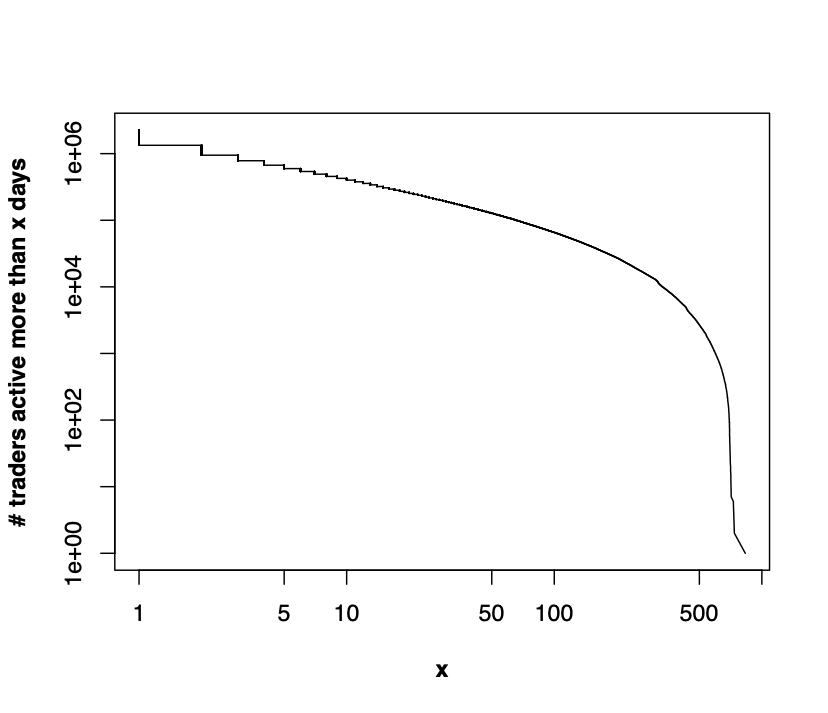}
        \includegraphics[width=0.45\textwidth,angle=0]{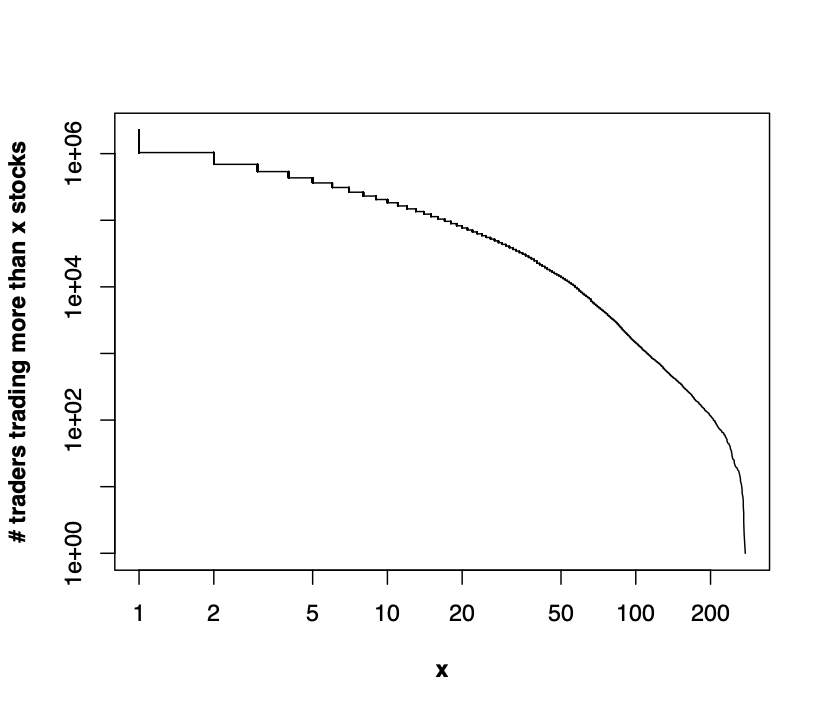}
	\caption{Non normalized complementary cumulative distribution function of the number of trading days when an investor was active (left panel) and of the number of stocks traded in the whole period (right panel). The axes in both panels are in double logarithmic scale. }\label{fig:dist}
\end{center}
\end{figure}

The analysis of the trading of individual investors in Italian stocks in the period January 2019-September 2021 is based on transaction reports collected by Consob for the Italian stocks, according to directive \href{https://eur-lex.europa.eu/legal-content/IT/ALL/?uri=CELEX:32004L0039&qid=1435044997184}{2014/65} by European Union, also called MiFID II, and the subsequent MiFIR.\footnote{In a nutshell, the MiFIDII/MiFIR regime has introduced new regulations for European financial markets and, among them, the transaction reporting obligation that requires investment firms or intermediaries executing transactions in financial instruments to communicate ``complete and accurate details of such transactions to the competent authority as quickly as possible, and no later than the close of the following working day”.} The relevant dataset was built aggregating the daily transactions of all investors operating in any of the Italian stocks, in the period from January 1, 2019 to September 30, 2021. In details, the dataset was built according to the following rules: i) all the information related to the identity of individual investors have been anonymized; ii) with reference to each stock (identified by its ISIN code), each data point keeps a record of: 
\begin{enumerate}
    \item anonymous identifier of the investor;
    \item type of investors (household, investment firm, non-MiFID firm, legal subject);
    \item trading venue of the operation (Borsa Italiana - MTA, London Stock Exchange - LSE, off-exchange, etc.);
    \item day of the operation;
    \item buy and sell volumes (in shares);
    \item buy and sell Euro volumes;
    \item number of buy and sell contracts;
    \item price of both the first and the last contracts (if there are more than one contract, otherwise they coincide);
    \item minimum and max prices of contracts (if there are more than one contract, otherwise they coincide);
    \item average price of buy (sell) contracts.
\end{enumerate}
In the period covered by the dataset, $2,253,707$ investors were observed, operating in 286 Italian stocks. As observed in several empirical studies of other markets \cite{svn}, investors are characterized by an extreme heterogeneity of trading frequency. This is the case also for the Italian stock market. The left panel of Figure \ref{fig:dist} shows the non normalized complementary cumulative distribution function (which provides the number of investors trading at least $x$ days as a function of $x$) of the number of trading days when an investor was active. We immediately see a very broad distribution (notice that both axes are in logarithmic scale) with a fat tail truncated by the maximal number of days of the dataset ($1003$ days). $41\%$ of the investors traded only one day and in Section \ref{sec:entryexit} we will provide some explanations for this. The heterogeneity of investors is also very significant when considering the number of different stocks traded by each investor in the whole period. The right panel of Figure \ref{fig:dist} shows the non normalized complementary cumulative distribution function of the number of stocks and again a large dispersion is observed with $54\%$ of the investors trading only one stock in the whole period.

The dataset contains some information on each investor that have been taken into consideration for the purpose of the analysis. First, five different categories are present, namely  {\it Persona Fisica}, {\it Conto Cointestato}, {\it Persona Giuridica} (legal entity), {\it Non-MiFID firm}, {\it Investment Firm}. The first includes individual households, while the second contains joint accounts of several households. The definition of {\it Retail investors} covers these two categories. For the households (individual and participating to a joint account), information on the date of birth, the gender, and the province of birth were analysed. More precisely, this information is fully available for (the vast majority of) Italian investors, while it is not available for foreigners, for which the information of the country of residence was considered. The last three categories are institutions and companies and, in the following, will be aggregated and indicated with the common name of {\it Firms}.

\section{How investors modified their trading during the lockdown}\label{sec:modify}

\begin{figure}
\begin{center}
        \includegraphics[width=0.42\textwidth,angle=0]{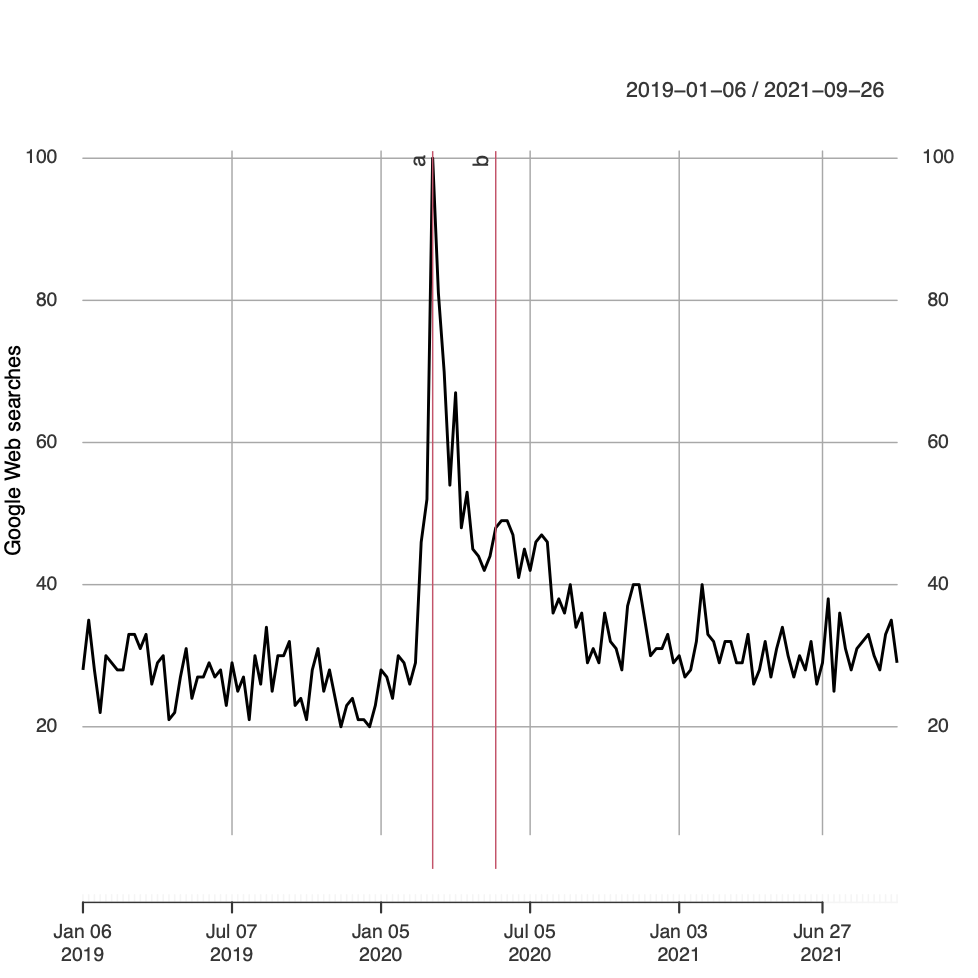}
        \includegraphics[width=0.5\textwidth,angle=0]{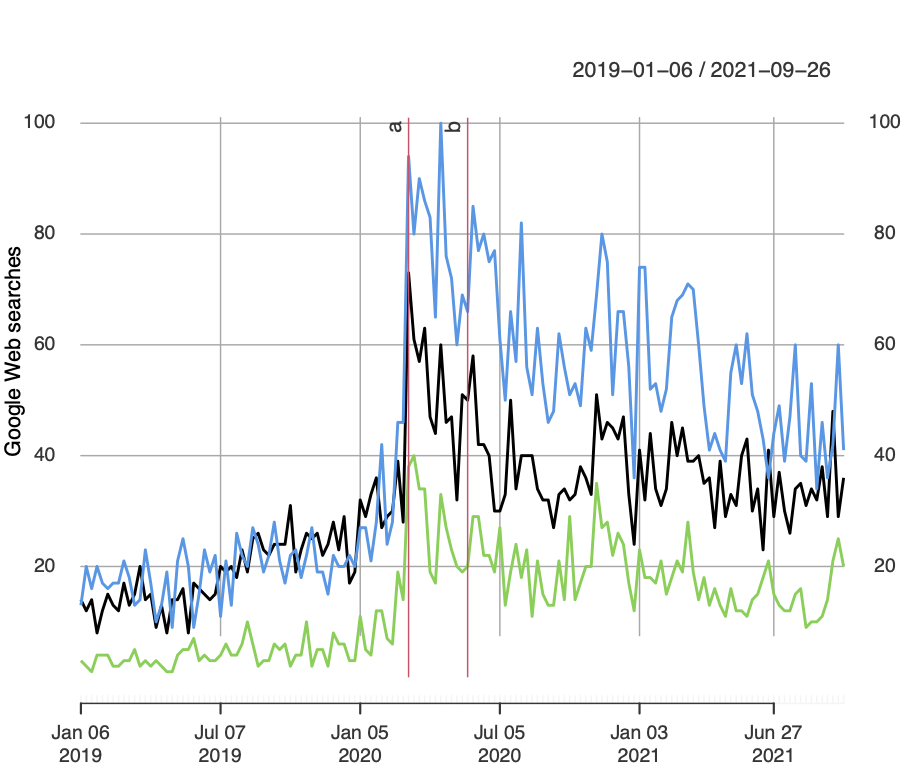}
	\caption{Left. Normalized volume of queries of the term ``Borsa di Milano" to Google from Italy in the investigated period (January 2019-September 2021). Right.  Normalized volume of queries of the term ``ENI titolo" (blue) ``ENI azione" (black), and ``ENI stock" (green) to Google from Italy in the investigated period (January 2019-September 2021). The red vertical lines identify the lockdown period.}\label{fig:google}
\end{center}
\end{figure}

To motivate our research, before investigating the actual behaviour of investors during the lockdown, a study about the popularity of stock markets in Italy was performed simply by using the number of times a specific query (``Borsa di Milano") has been searched on Google in Italy. As it is known, this information is available via Google Trends (\href{https://trends.google.com/trends/}{https://trends.google.com/trends/}) on a weekly basis and renormalized in such a way that the maximum over the period is arbitrarily set equal to $100$. The time series of the number of such queries is shown in the left panel of Figure \ref{fig:google}, while the right panel shows the dynamics of the number of times three queries related to a specific stock (ENI). For the first case  we immediately observe a more than threefold increase of the number of queries at the time of the lockdown, indicating a sudden surge of interest on the stock market when the government imposed the lockdown. A similar increase is observed for the queries related to ENI. As we will see below, this increase in Internet searches is mirrored by a similar increase in trading activity of households and by the arrival of a population of new households investing in the market for the first time. We will also show some analyses indicating a change in the trading strategy of households when lockdown was put in place. From the figure it is possible to note also that the number of searches after the lockdown is on average larger than the one before it, indicating some sort of persistent shift toward a new regime. This might be due to a change in the population of investors, likely triggered by the lockdown, as it will be documented below.

\subsection{A case study: ENI}

The first analysis concerns a specific, yet representative, stock, ENI, which is the second largest publicly quoted company by market capitalization. The analysis aims at studying how, in aggregate, retail investors and firms modified their trading behaviour during the investigated period. The first quantities analysed are the total traded volume and the net imbalance, defined as the purchased volume minus the sold volume. Both quantities are in Euro. The use of Euro instead of shares allows later to aggregate different stocks traded in the market. 

\subsubsection{Trading volume}

\begin{figure}
\begin{center}
        \includegraphics[width=0.4\textwidth]{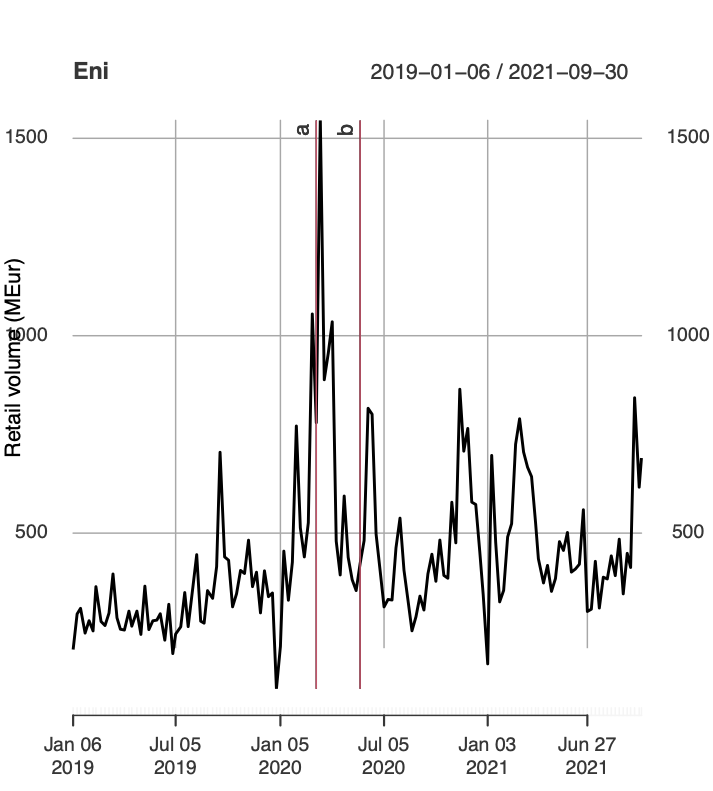}
	\includegraphics[width=0.4\textwidth]{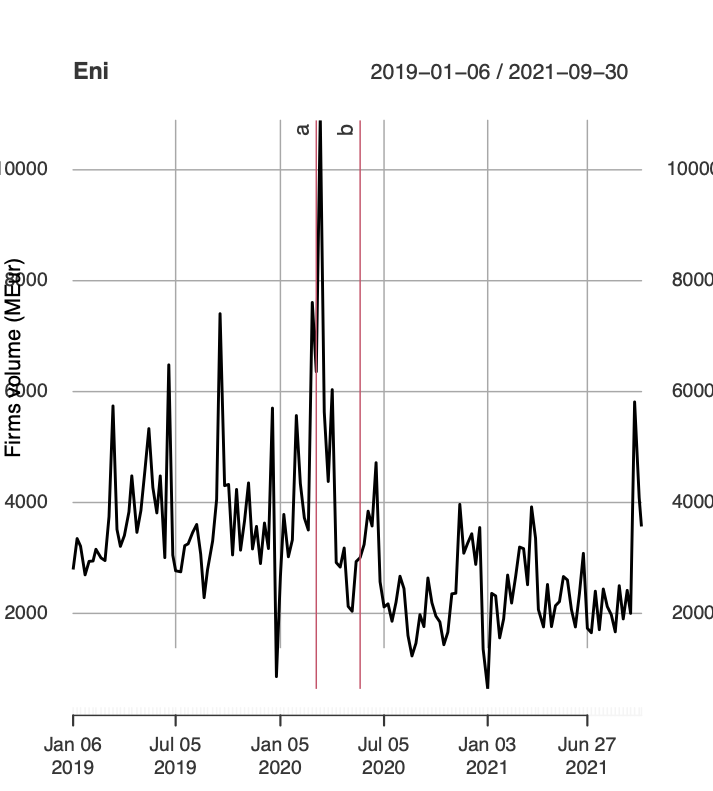}
	\caption{Total volume of ENI traded weekly by retail investors (left panel) and firms (right panel). The two vertical lines indicate the first lockdown.}\label{fig:enivolume}
\end{center}
\end{figure}

\begin{figure}
\begin{center}
 \includegraphics[width=0.4\textwidth]{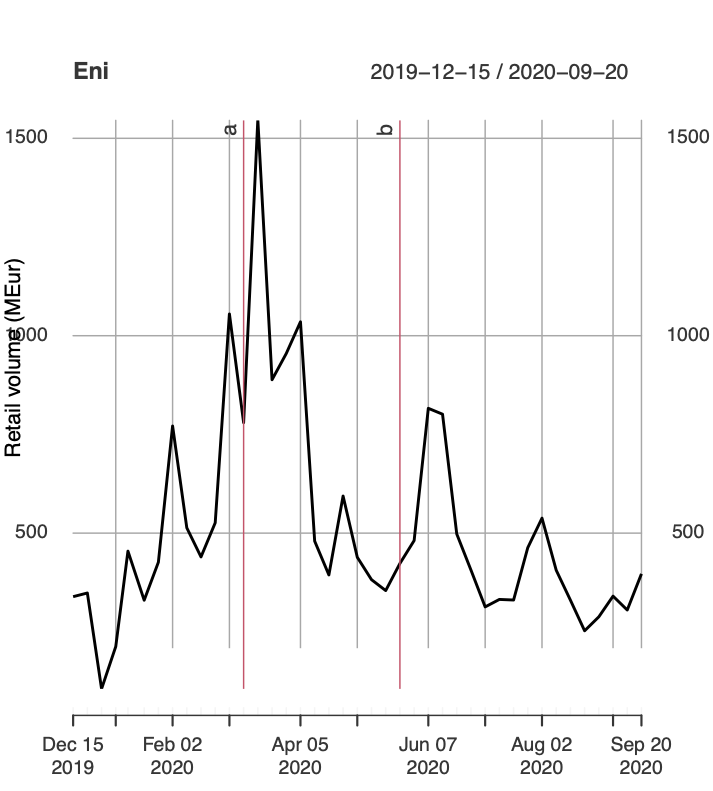}
\end{center}
\caption{Total volume of ENI traded weekly by retail investors around the first lockdown (vertical lines).}\label{fig:enivolumezoom}
\end{figure}

Figure \ref{fig:enivolume} shows the weekly time series of the traded volume of ENI by retail investors (left panel) and firms (right panel).
The increase in the trading volume for both categories of investors is clear. The former group increases the volume from $\sim 300 M$Eur to more than $1,500 M$Eur per week.
 A smaller but still very significant increase is observed for firm's volume.  It is important to notice that  the increase in volume started  slightly before the lockdown (see Figure \ref{fig:enivolumezoom}) and the increase is very localized in the first 2-3 weeks. As discussed above, the first localized restrictions were in place from the last week of February 2020. For ENI in part this might be due to the price dynamics (see the left panel of Figure \ref{fig:contrarian} below), and in fact for the aggregate market we will see that the shift is visible only one week before the lockdown. Also, the dynamics of Google queries related to ENI (right panel of Figure \ref{fig:google}) shows an increased interest toward this stock before the lockdown. As we will see in Section \ref{sec:new}, the increase in volume is partly due to the arrival of new investors. 

\subsubsection{Net Trading}

\begin{figure}
\begin{center}	
	\includegraphics[width=0.35\textwidth]{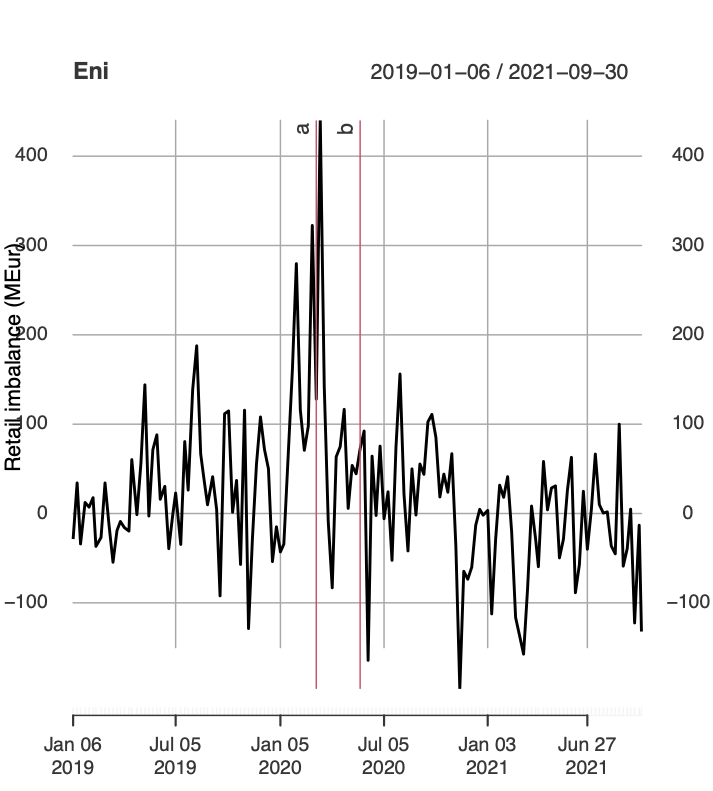}
	\includegraphics[width=0.35\textwidth]{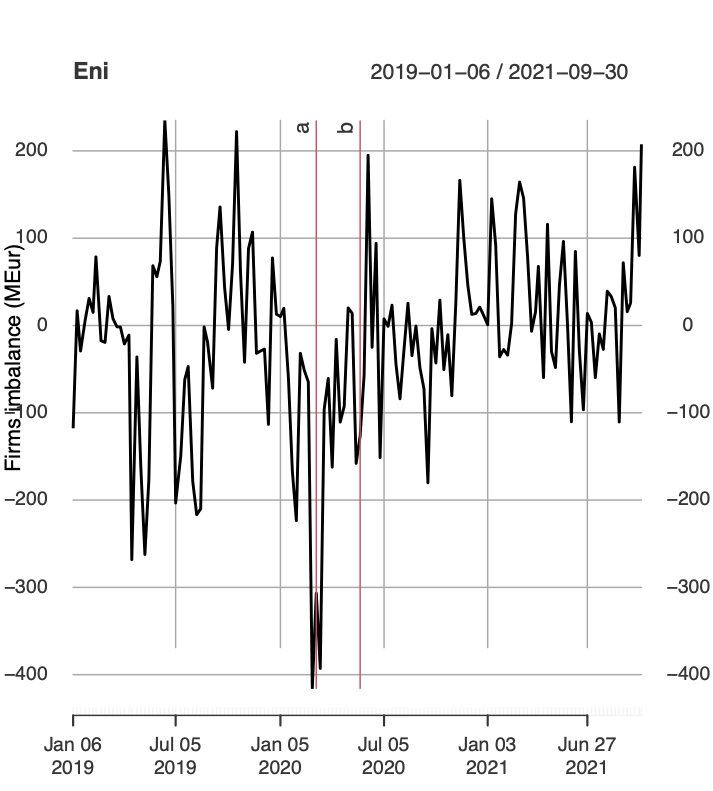}
	\caption{Net trading of ENI traded weekly by retail investors (left panel) and firms (right panel). The two vertical lines indicate the first lockdown.}\label{fig:enisignedvolume}
\end{center}
\end{figure}

We now consider the {\it Net Trading} $NT_{i,t}$, defined, for a specific group of investors (e.g. households, firms, etc), as 
$$
NT_{i,t}= \emph{Buy Euro Volume}_{i,t}-\emph{Sell Euro Volume}_{i,t}
$$
where $i$ indicates the stock and $t$ the considered time. A normalized version of $NT$ (for individuals) has been introduced in \cite{Kaniel}. The choice of Euro instead of shares has been done to be able later to aggregate more $NT_{i,t}$ for more stocks.
 Figure \ref{fig:enisignedvolume} shows the weekly time series of the net trading of ENI by retail investors (left panel) and firms (right panel). Clearly the two quantities should be one the opposite of the others if the database contains no errors, since the total net trading is by definition equal to zero. The comparison between the two panels shows that this is approximately true, thus from now on we will consider only the net trading of retail traders.

It is possible to observe that, while before the pandemic the net trading, both of firms and of retail traders, typically oscillates around zero, a dramatic difference is observed around the lockdown. From a month before the lockdown to a couple of weeks inside it the retail traders became net buyers while firms became net sellers. As we will see below, when considering all stocks this behaviour is confirmed and the peak more localized inside the first part of the lockdown. Notice however that, as shown in Figure \ref{fig:stringency}, the restrictions in Italy started a couple of weeks before the lockdown.

\subsection{All stocks}

\begin{figure}
\begin{center}	
	\includegraphics[width=0.4\textwidth]{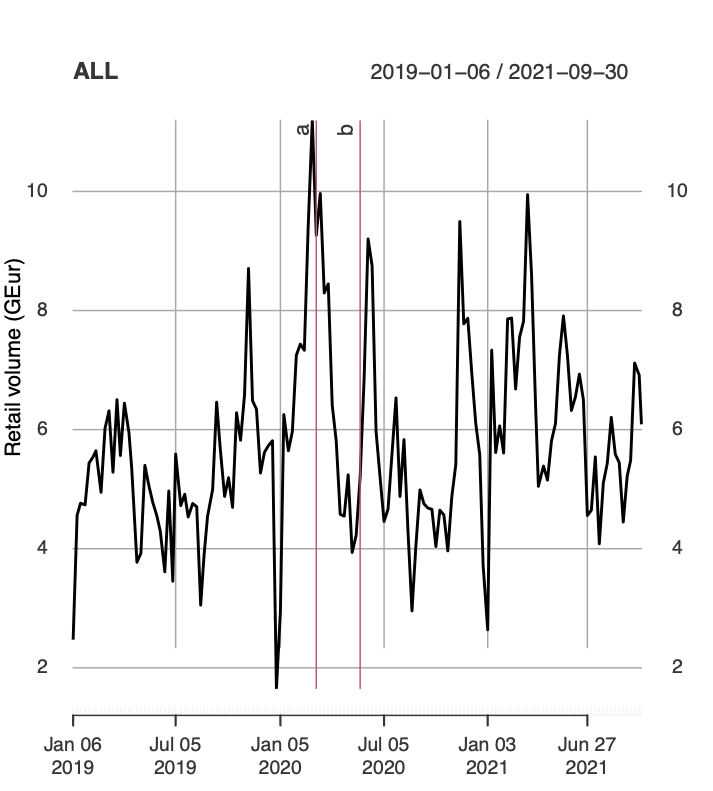}
	\includegraphics[width=0.4\textwidth]{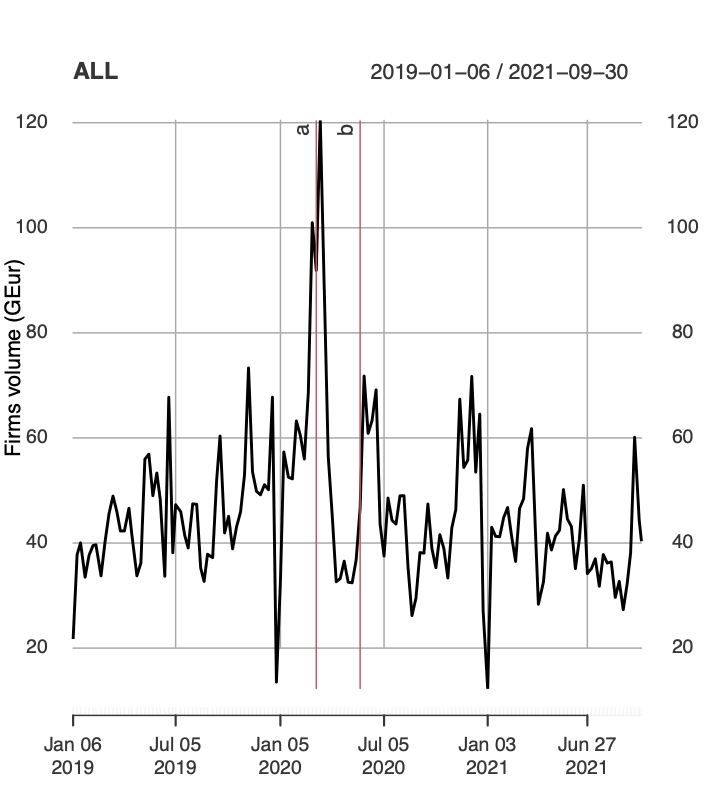}
	\caption{Total volume  traded weekly by retail investors (left panel) and firms (right panel) in all companies. Volume is in Euro. The two vertical lines indicate the first lockdown.}\label{fig:allvolume}
\end{center}
\end{figure}

In the second step all the stocks were considered. Figure \ref{fig:allvolume} shows the time series of the trading volume of households and firms during the investigated period. As for ENI, it is possible to observe an abnormal increase of trading volume for both categories of investors starting a couple of weeks before the lockdown and reaching the peak at its beginning.

\begin{figure}
\begin{center}	
	\includegraphics[width=0.4\textwidth]{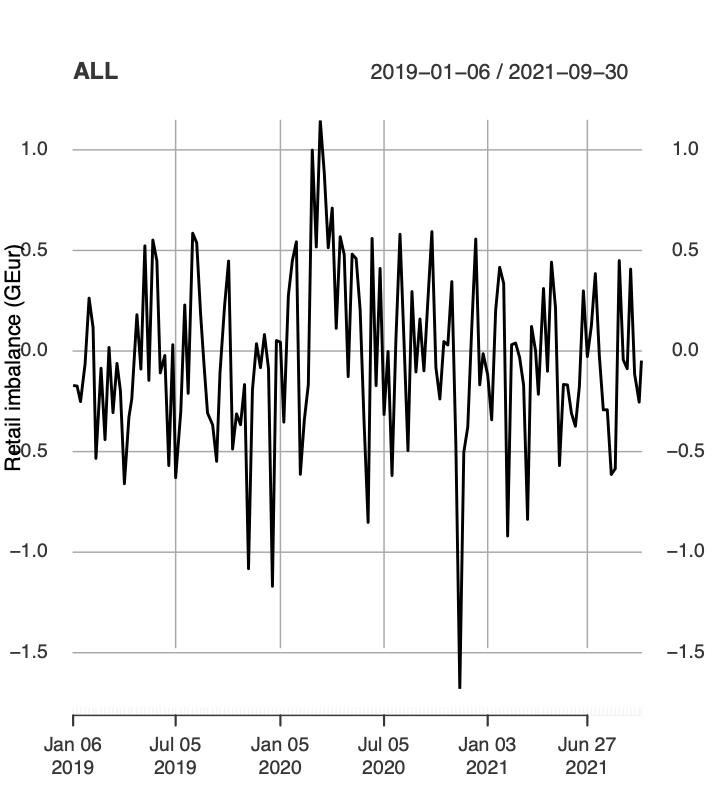}
	\includegraphics[width=0.4\textwidth]{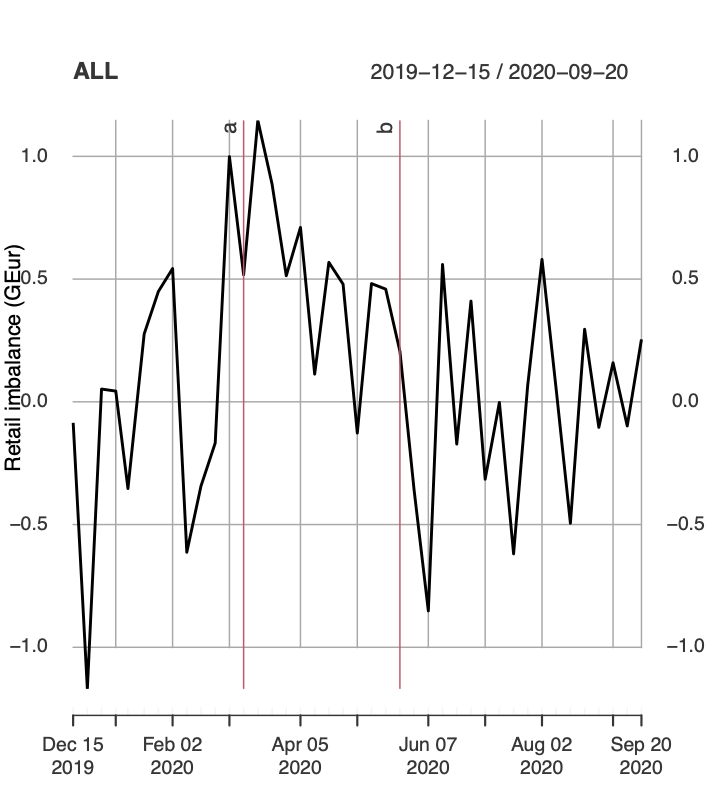}
\end{center}
\caption{Net trading (in Euro) by retail investors in the investigated period (left panel ) and a zoom around the lockdown (right panel) indicated by the red vertical lines.}\label{fig:allimbalance}
\end{figure}

Also, the behaviour of net trading for the pooled set of all the stocks, i.e.
$$
NT_{t}=\sum_{i=1}^{N_{stocks}}N_{i,t}
$$
is similar to what was observed for ENI (see Figure \ref{fig:allimbalance} where we show the results only for households given that the result for firms is very close to the opposite). The two panels clearly indicate that households became net buyers during the lockdown (and possibly one-two weeks before), while in the other periods their net trading oscillates around zero. In the next subsection we investigate whether such strong polarization of households could be explained as a reaction to concurrent and past price dynamics.

\subsection{The role of price return and contrarian behaviour of households}

\begin{figure}
\begin{center}	
    \includegraphics[width=0.4\textwidth]{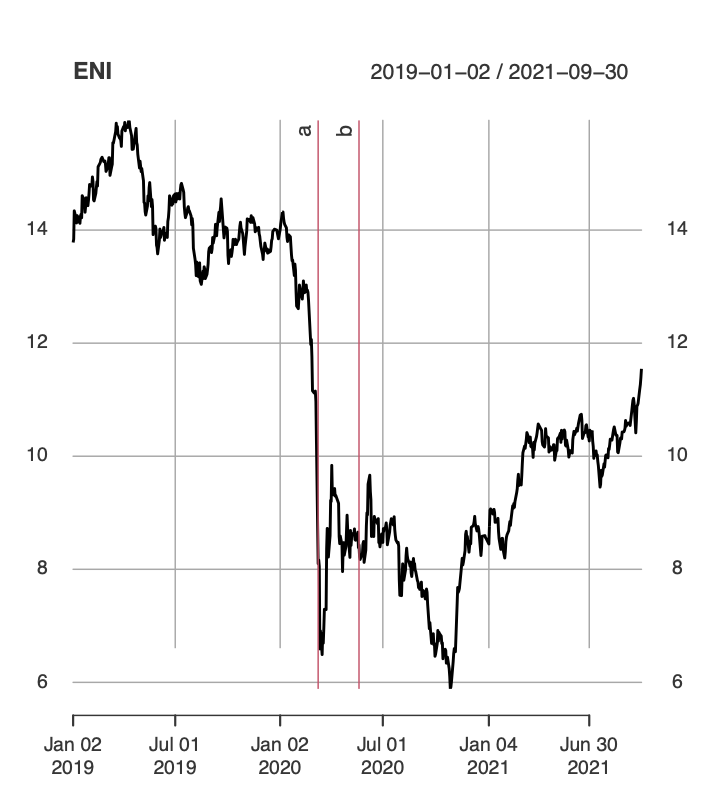}
    \includegraphics[width=0.4\textwidth]{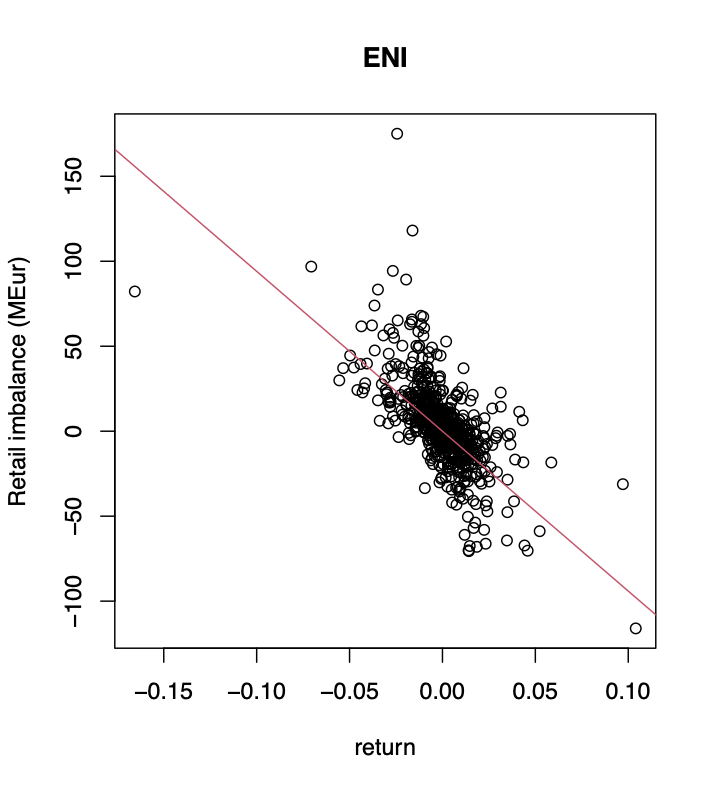}
\end{center}
\caption{Left panel. Time series of ENI closing price. The two vertical lines indicate the first lockdown. Right panel. Scatter plot between net trading of retail investors and close to close return. The red line is the best linear fit (each point is a trading day).  }\label{fig:contrarian}
\end{figure}

One of the possible origins of the change in trading behaviour and the net buying of retail investors is the well-known contrarian behaviour of households which has been empirically shown in several papers (e.g. see \cite{Kaniel,pagano}). A contrarian strategy buys when the price declines and sells when the price increases. It can be interpreted as a strategy that bets on the mean reversion of prices and interpret significant price changes as transient displacement from an equilibrium price, due either to overreaction or to liquidity effects. The contrarian behavior can be identified considering the correlation of net trading either with contemporaneous price returns or with past price changes. In our case, the daily resolution of the data does not allow to discriminate intraday contrarian behavior, which can be evidenced by looking at synchronous correlations.

The potential relevance of contrarian behavior in the investigated period can be understood by looking at the price dynamics during the Covid pandemic. As a representative example, the left panel of Figure \ref{fig:contrarian}
shows the close price time series of ENI in the investigated period. It is evident that the price of ENI (as those of the other stocks) displayed a large decline and this, because of the contrarian nature of households, might have triggered both the increase in retail volume and their net buying behaviour.

To investigate whether a contrarian behavior of household trading is observed in our dataset, the right panel of Figure \ref{fig:contrarian} shows a scatter plot between the net trading by retail investors and close price returns over the whole investigated period. There  is a clear  negative dependence between the two variables: the Pearson correlation coefficient is $-0.65$ and the coefficient of determination of the regression is $R^2=0.42$. This indicates a strongly statistically significant negative dependence. In aggregate retail investors buy when the price drops and sell when the price increases.

\begin{figure}
\begin{center}	
	\includegraphics[width=0.45\textwidth]{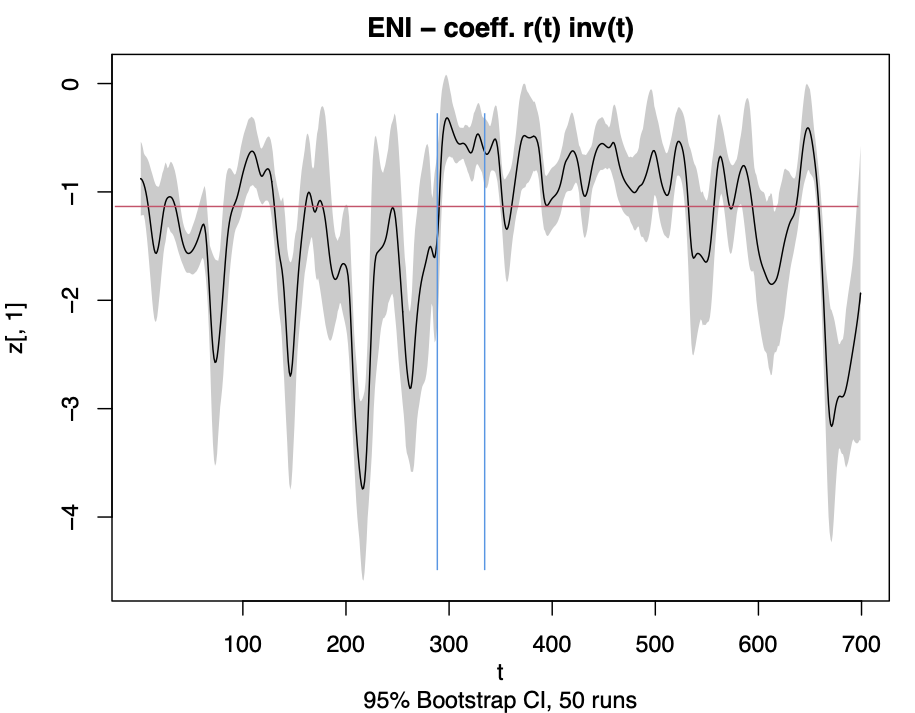}
\end{center}	
\caption{Coefficient $\beta_{i,t}$ of the local regression model of Eq. \ref{eq:invret} describing the synchronous relation between net trading of households and  daily return of stock $i$. The considered stock is ENI. The grey region is the $95\%$ confidence interval obtained with bootstrap. The blue vertical lines indicate the lockdown period, while the red horizontal line is the value of the constant coefficient $\beta_i$ in a standard  OLS regression.}\label{fig:tvreg1}
\end{figure}

The contrarian behaviour could explain why households were net buyer during the lockdown, but is this explanation correct or can we find evidence of a change in strategy of the households during lockdown? In order to answer this question, we use a machine learning kernel regression method to estimate the local dependence between return and net trading and estimate its dynamics. Specifically, the time-varying regression between daily return and net trading by household of asset $i$ is
\begin{equation}\label{eq:invret}
NT_{i,t}=\alpha_{i,t}+\beta_{i,t} ret_{i,t}+\epsilon_{i,t}
\end{equation}
where $ret_{i,t}$ is the daily stock return (close-to-open) of asset $i$, while $\epsilon_{i,t}$ is a residual term. The coefficients $\alpha_{i,t}$ and $\beta_{i,t}$ are time-varying and are estimated locally. Specifically, this time-varying regression model is estimated by using the R package tvReg \cite{tvreg}, which uses a kernel method with an optimally chosen bandwidth. 
By considering the stock ENI,
Figure \ref{fig:tvreg1} shows the estimated time-varying $\beta_{i,t}$ together with the 95\% confidence bands. We observe that for the whole period the coefficient is negative, indicating a time consistency in the contrarian behaviour of retail investors. However, what it is more evident from the figure is the sharp increase of $\beta_{i,t}$ when the lockdown started and the fact that this coefficient remains less negative for several months. This clearly indicates that household investors became {\it less} contrarian during and after the lockdown, i.e. they bought less (or sold more) than expected by looking at their trading behaviour before the lockdown. This might be another indication of a structural change in the household investor population at the time of the lockdown. Solid indications that this change might be driven by a change in the composition of household investors are provided below.

\medskip

\begin{figure}
\begin{center}	
	\includegraphics[width=0.32\textwidth]{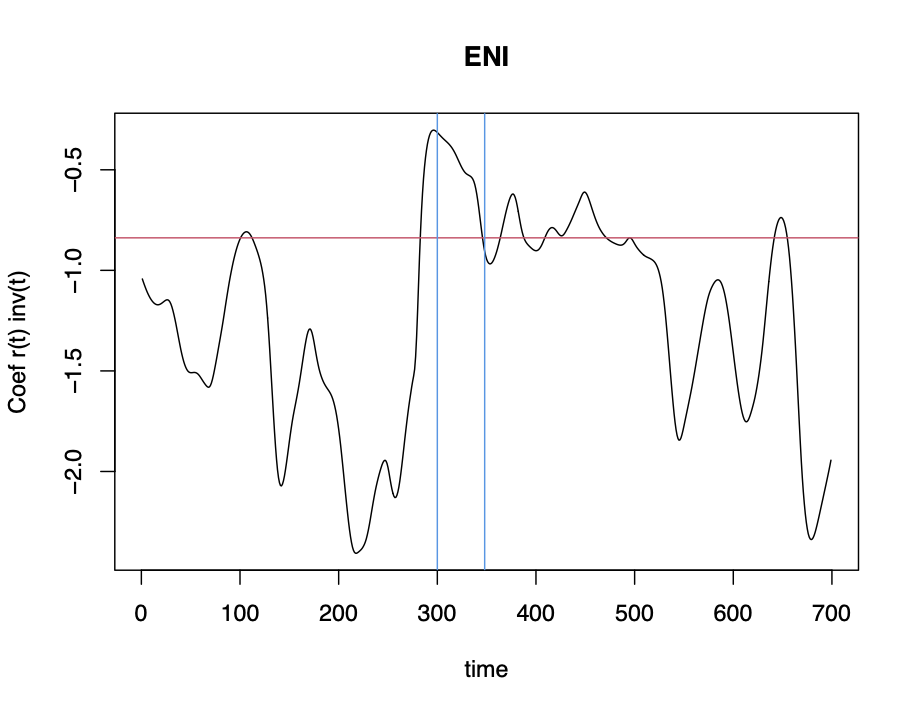}
	\includegraphics[width=0.32\textwidth]{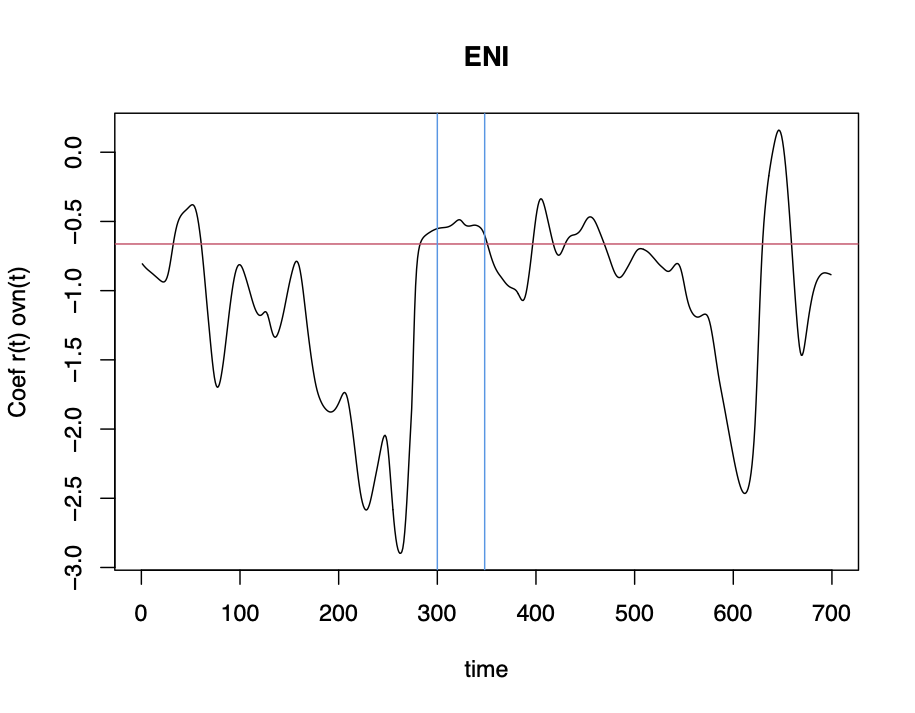}
	\includegraphics[width=0.32\textwidth]{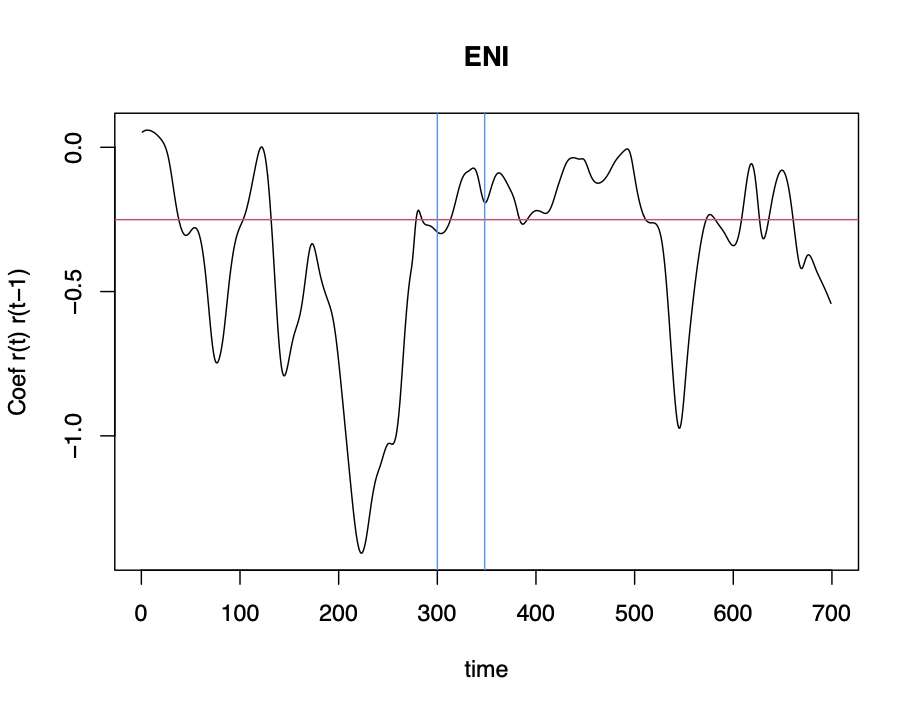}
\end{center}
\caption{Coefficients $\beta^1_{i,t}$ (left), $\beta^2_{i,t}$ (middle), and $\beta^3_{i,t}$ (right)   of the local regression model of Eq. \ref{eq:invret2} describing the relation between signed volume of households and simultaneous daily return, past overnight return, and past daily return, respectively. The considered stock is ENI. The blue vertical lines indicate the lockdown period, while the red horizontal line is the value of the constant coefficients $\beta_i$s in a standard multivariate OLS regression}\label{fig:tvreg2}
\end{figure}

As a robustness control and also to avoid the use of regression only of simultaneous variables, we consider a richer time-varying regression model which includes as regressors also past returns. More specifically, we consider the model 
\begin{equation}\label{eq:invret2}
NT_{i,t}=\alpha_{i,t}+\beta^1_{i,t} ret_{i,t}+\beta^2_{i,t} ovn_{i,t-1}+\beta^3_{i,t} ret_{i,t-1}+\epsilon_{i,t}
\end{equation}
where $ovn_{i, t-1}$ is the overnight return between day $t-1$ and $t$ and $ret_{i,t-1}$ is the close-to-open price return of the previous day for stock $i$. These new regressors are meant to capture a genuine contrarian behaviour, which is generically defined as a strategy which sells (buys) when {\it past} returns are positive (negative). Figure \ref{fig:tvreg2} shows the time series of the estimated coefficients. In all cases they are (almost always) negative, reinforcing the contrarian interpretation of the household aggregate trading activity. More interestingly, at the time when the lockdown was put in place there is a clear upward shift of all the coefficients, similarly to the simpler model of Eq. \ref{eq:invret}, and indicating a weaker contrarian behaviour.

In conclusion, the analysis of the net trading of households showed that an abrupt change occurred when the first restriction due to Covid-19 pandemic were put in place: household traded more, they were net buyers, but this is not explained by their usual contrarian behaviour. In fact, the time-varying regression showed that a sharp change in the correlation between household net trading and contemporaneous and past return occurred at the beginning of the lockdown and, at least in some cases, appeared to remain also after its end. The next section will show that the main reason for this change was a significant modification of the composition of household population.

\section{The new investors}\label{sec:new}

\subsection{Entry-exit dynamics of investors}\label{sec:entryexit}
It is interesting to investigate how the composition of household group changed during and after the lockdown. To this end we divide the whole investigated period in three subsamples: 
\begin{enumerate}
\item the pre-lockdown period lasting from January 1, 2019 to March 8, 2020 (433 calendar days);
\item the lockdown period lasting from March 9, 2020 to May 18, 2020 (71 calendar days);
\item  the post-lockdown period lasting from May 19, 2020 to September 30, 2021 (500 calendar days).
\end{enumerate}

As mentioned in Section \ref{sec:data}, in total there are $2,253,707$ unique investors (including both households and firms). Considering all the stocks, in the pre-lockdown period, $1,151,234$ unique investors traded at least once. These investors are called {\it prelock} investors. When considering the lockdown period (March 9 to May 19, 2020)  we find that $563,575$ unique investors traded at least once. Surprisingly, {\it of these $184,987$ ($32\%$) had never traded in the pre-lockdown period.} These investors are defined {\it newcomers}. It is important to clarify that, given the nature of our data, it is not possible to tell with certainty if these investors really entered for the first time in the Italian stock market during the relatively short time period of the lockdown. In fact, the dataset tracks the trading activity from January 2019. However, given the relatively long pre-lockdown period at our disposal (January 2019 - March 2020), it is very likely that for a large fraction of these  $184,987$ investors the lockdown period was the first occasion to enter the Italian stock market. Moreover, we will show below (Section \ref{sec:demo}) that the demographic characteristics of these investors are markedly different from those in the pre-lockdown period, strongly suggesting that they were truly newcomers to the financial market.

\begin{figure}
\begin{center}	
	\includegraphics[width=0.4\textwidth]{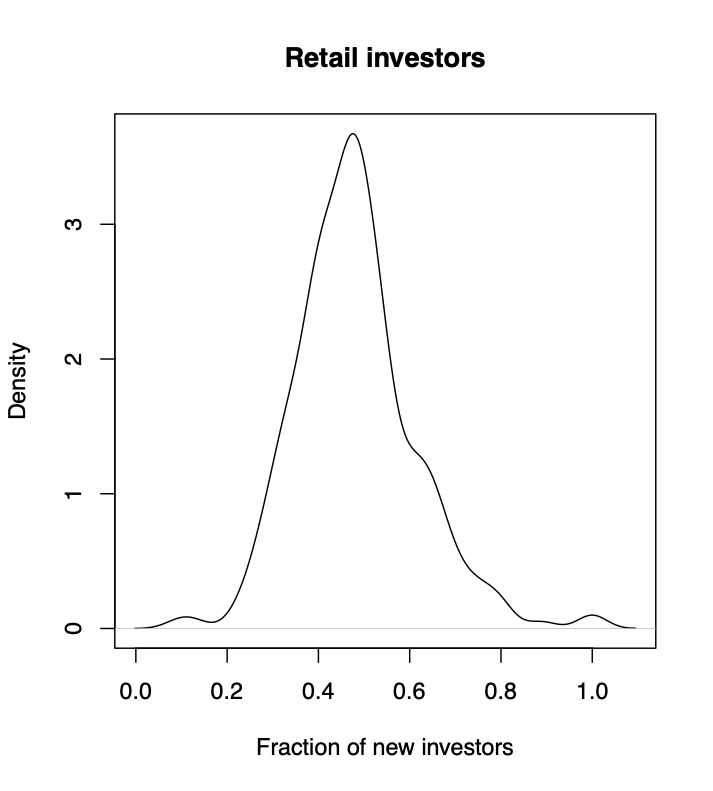}
\end{center}
\caption{Density estimation of the distribution of the fraction of investors who traded for the first time a given stock during the lockdown period.}\label{fig:fractionnew}
\end{figure}

Repeating the same analysis for each individual stock, we compute the fraction of investors who traded for the first time a given stock during the lockdown\footnote{Once more, we do not know if an investor traded an asset before January 1, 2019, thus it is more precise to say that they have never traded the stock in the period January 1, 2019 - March 8, 2020.} Figure \ref{fig:fractionnew} shows the distribution of this fraction for the $286$ Italian stocks. The distribution is centered around $50\%$ indicating that, on average, half of the investors trading a given stock during the lockdown had never traded it in the pre-lockdown period\footnote{In this analysis it is possible that a retail investor who traded stock A for the first time during the lockdown had traded stock B in the pre-lockdown period, thus this investor is not a newcomer according to the definition given before, but it will be considered in the fraction of new investors in Figure \ref{fig:fractionnew}.} This is a further indication that the trading strategies changed significantly during the lockdown, not only because of the arrival of newcomers in the whole market but also because there have been many new investors when restricting the analysis to each individual stock.

As a next step we divided the sample in two subsets, one composed by those who traded only in one day and those who traded more than one day (any stock).  We find that the first sample is quite large, being composed by $919,418$ (41\%) investors who traded only in one day. 
An in-depth investigation of this subset indicates that the average number of such investors is roughly constant but there are very large peaks corresponding to special events such as Public Exchange Offers. Thus these episodic investors, who trade only one day in the whole period, might have non strategic reasons for trading. For this reason, from now on the attention is focused on those investors who traded at least two days in the investigated period.

\begin{figure}
\begin{center}	
	\includegraphics[width=0.4\textwidth]{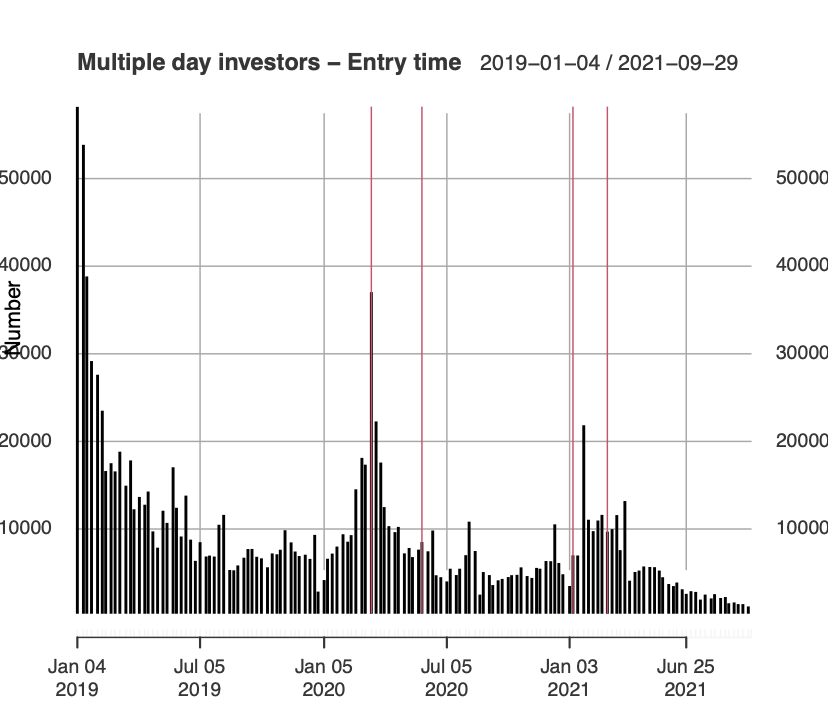}
	\includegraphics[width=0.4\textwidth]{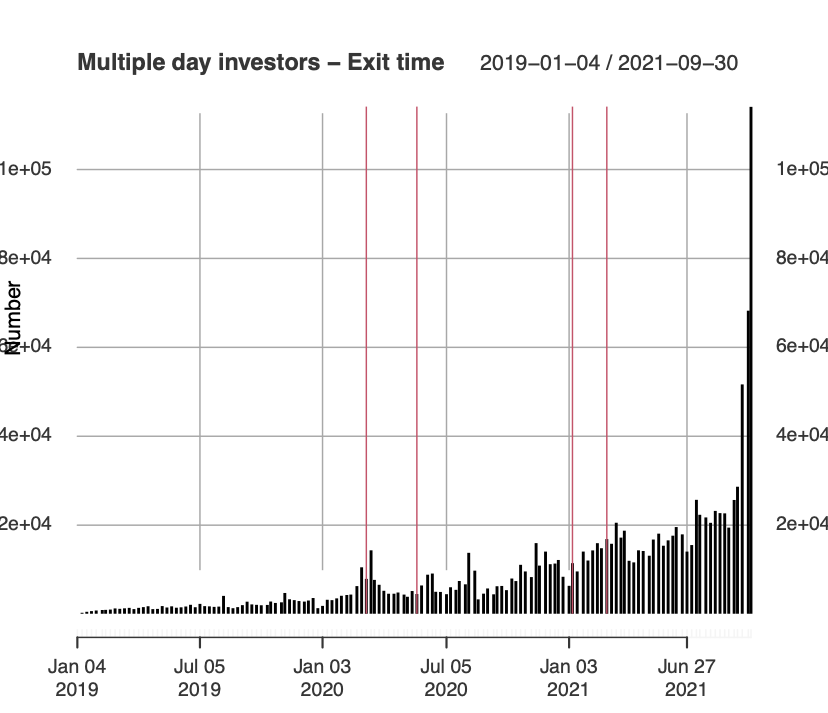}
\end{center}
\caption{Number of (non single day) investors who made the first (left panel) and last trade (right panel) in each week of the investigated period. The vertical lines are the two lockdown periods.}\label{fig:entryexit}
\end{figure}

In the database there are  $1,334,289$ investors who traded at least two days in the whole period of analysis and we study when they made the first and the last trade in order to quantify dynamically their entry-exit activity in the Italian stock market. Figure \ref{fig:entryexit} shows the number of (non single day) investors who made the first (left panel) and last trade (right panel) in each week of the investigated period. Clearly the large values observed on the left of the left panel and on the right of the right panel are an artefact of the dataset, since, by definition, a large number of investors made the first trade at the beginning of the investigated period and the last one toward the end. Much more interesting are the two clear peaks in the number of entries observed around the first and the second lockdown (left panel). This indicates that the lockdowns triggered the interest for a large number of new investors who had not traded before. Interestingly the pattern in the left panel during the first lockdown mirrors the peak in Google searches displayed in Figure \ref{fig:google}.

Of the 788k prelock investors (trading at least two days), $24\%$ did not traded anymore after the end of the lockdown. Even if it is difficult to know whether they truly exited from the market or simply held their position until the end of our dataset (September 2021), the large percentage suggests that a significant fraction was indeed convinced to stop trading or slow down after the abrupt transition brought by Covid lockdown. The P\&L analysis shown in Section \ref{sec:pl} supports this hypothesis.
A quite small fraction of the newcomers (19k investors) traded only during the lockdown, suggesting an opportunistic trading behavior possibly related to entertainment seeking or gambling. The other 124k newcomers (trading at least two days) continued to trade after the end of the lockdown, suggesting a persisting presence in the stock market, triggered by the lockdown.
Finally, a relatively large number of investors (404k) made the first and last trade after the lockdown, constituting a new population of investors numerically comparable to the prelock investors.

\subsection{Demographic composition of the new investors}\label{sec:demo}

As observed in the previous section, during the first lockdown a large number of new investors started to trade Italian stocks. But are they demographically different from those trading before the lockdown? The availability of electronic trading platform available from home and the lack of clear separation between work time and free time likely allowed households to consider the possibility of trading during the day. 

Considering only investors which traded more than one day in the whole sample, we then restrict our attention to households, thus discarding {\it Conti Aggregati} (and of course Firms). We also considered only Italian investors, since only for them the demographic information is reliable. For each of them we identify the first day of our sample when they traded any  Italian stock and for each week we investigate  the age (year of birth) and the gender of the household investors entering the market for the first time in that week. 

\begin{figure}
\begin{center}	
	\includegraphics[width=0.4\textwidth]{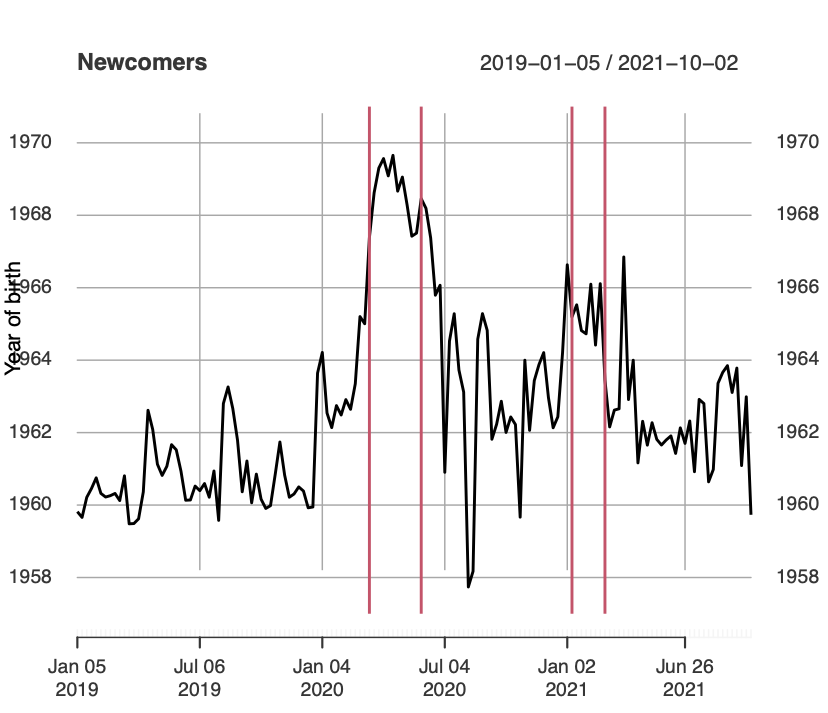}
	\includegraphics[width=0.4\textwidth]{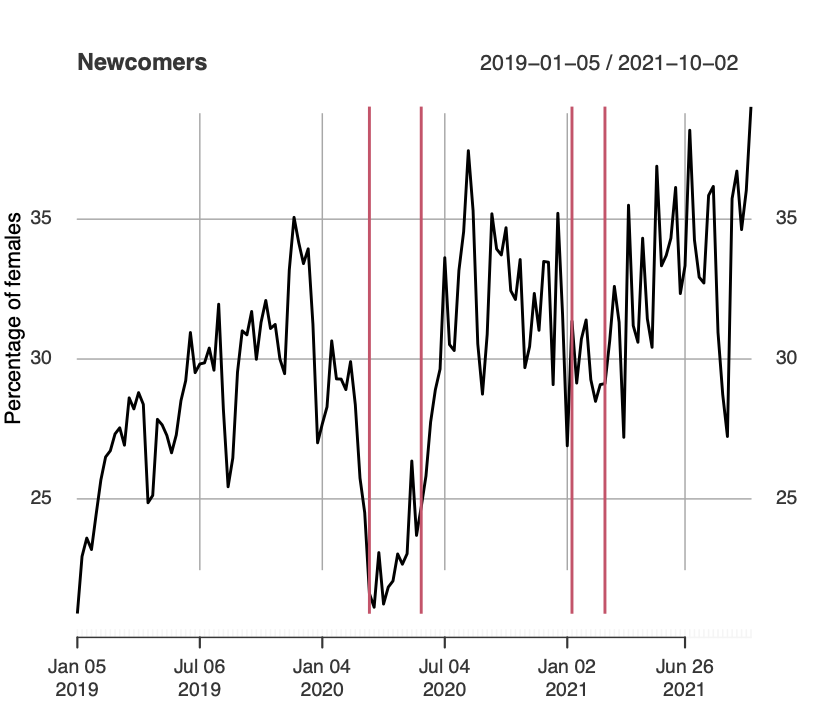} \\
\end{center}
\caption{Left panel. Average year of birth of the investors which traded for the first time (in our database) in each week. Right panel. Fraction of females among the  investors which traded for the first time (in our database) in each week. The vertical lines identify the lockdown periods.}\label{fig:demo}
\end{figure}

In the left panel of Figure \ref{fig:demo} we observe that the average year of birth of the newcomers before the lockdown is 1961 (corresponding to an average age of 59). It is evident from the left panel that during the lockdown the average year of birth raised to 1969-1970 (corresponding to an average age of 50-51). Notice also that the average year of birth of the newcomers after the lockdown went down, but on average it remained roughly five years larger than before the lockdown indicating that the new investors, also after the lockdown, tend to be younger than before. In other words, our result suggests that the lockdown had the merit of attracting more young investors toward the Italian stock market. Finally, it is interesting to notice that during the second lockdown (here defined as January 7, 2021- February 15, 2021) we observe another increase of the year of birth, suggesting another influx of young investors during this mobility restriction period.

Another abrupt change in the composition of the households trading for the first time is visible when studying their gender. The right panel of Figure \ref{fig:demo} shows the weekly fraction of females among the household investors trading for the first time as a function of the week when they enter the market. It is evident that, while this fraction is typically above 30\%, during the lockdown it dropped around 22\% and then bounced back when the period ended. Again, a similar drop in the fraction of females is observed in the second lockdown. Interestingly, after the first lockdown the fraction of females trading for the first time increases to levels higher than before.

As mentioned above, a significant fraction of newcomers, i.e. households entering the market during the first lockdown, made their last trade before the end of the lockdown (remember we are considering investors trading at least in two days). We term them as ``impatient" and those who made the last trade after the lockdown as ``patient". There is evidence that impatient investors were on average three years younger than the patient ones, with a mean year of birth of $1968.9$  and $1965.8$, respectively (the p-value of the t-test is smaller than $2.2  ~10^{-16}$). Considering their gender, the fraction of males is larger ($77\%$) for the impatient investors than for the patient ones ($73\%$). 

In conclusion, during the lockdown a large number of new investors entered the Italian stock market. Demographically they were quite different from the typical investors present before the lockdown, since they were characterized by the fact of being significantly younger and more frequently male. The relatively small number of investors trading only during the lockdown seems to reduce the importance of the gambling hypothesis, even if it this population has demographic characteristics which are different from those of the other newcomers.

\subsection{Are the newcomers more skilled investors?}\label{sec:pl}

\begin{figure}
\begin{center}	
	\includegraphics[width=0.49\textwidth,angle=0]{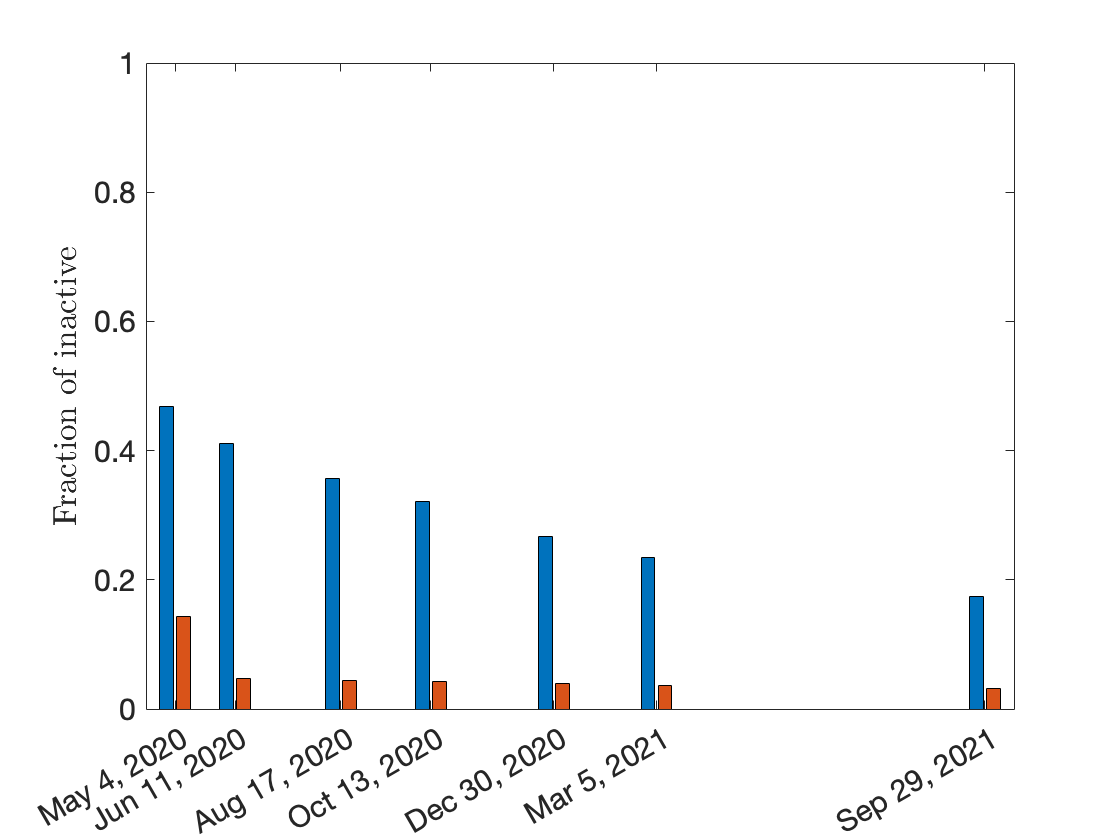}
	\includegraphics[width=0.49\textwidth,angle=0]{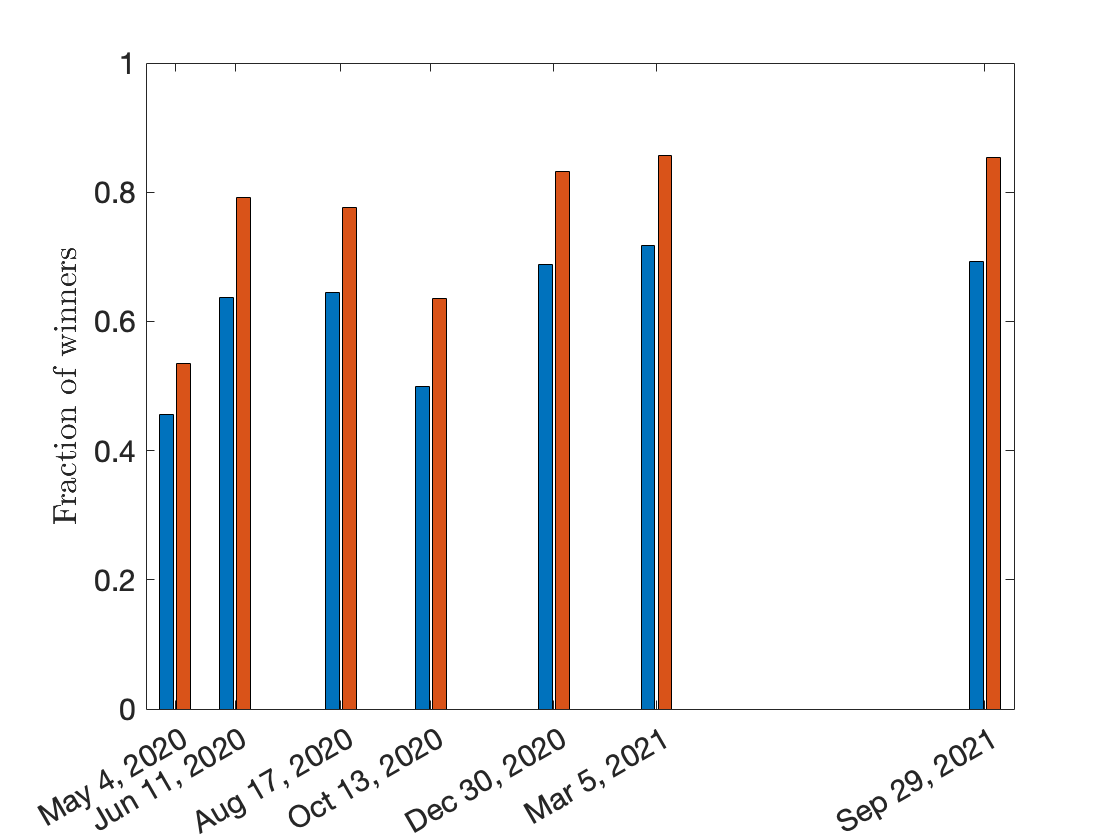}
\end{center}
\caption{Left panel. Fraction of inactive investors, namely those with zero P\&L, at the seven considered dates. Right panel. Fraction of investors (over the active ones) with positive P\&L at the seven considered dates. The blue bars represent the prelock investors, while the red one the newcomers.}\label{fig:pl}
\end{figure}

Did the newcomers make a profit from their trading? How are they skilled in comparison with prelock investors? As known, measuring profit and loss using the type of data at our disposal is typically very complicated. The first reason is that assessing transaction costs, fees, taxes is typically an hard task, without knowing the details of the trading procedure used by the investors. The second reason is related to the fact that we do not know the initial portfolio of the investors, hence we cannot evaluate their complete profit and loss (P\&L).  As done typically in the literature (see e.g. \cite{Barber}), we consider a subset of trades and we evaluate their potential P\&L marking-to-market the positions at specified dates. In particular, we are interested in the P\&L of the trades performed during the lockdown period and evaluated at a certain number of subsequent dates. We choose the following dates:

\begin{enumerate}
\item May 4, 2020 - End of the lockdown and beginning of ``Fase 1";
\item June 11, 2020 - End of ``Fase 1" and ``Fase 2" and beginning of ``Fase 3", corresponding to the opening of almost all commercial activities (summer period);
\item August 17, 2020 - End of ``Fase 3" (summer) and introduction of new restrictions, e.g. closure of discos;
\item October 13, 2020 - A number of administrative orders (DPCMs) reintroduce a severe anti-contagion policy, resulting in local lockdowns at a regional level depending on local epidemic indicators;
\item December 30, 2020 - End of the 2020 year;
\item March 5, 2021 - The first administrative order by the new prime minister Mario Draghi. The severe anti-contagion policy continues;
\item September 29, 2021 - The last day in the dataset.
\end{enumerate}

Since some of the stocks are very illiquid, we focus on the $214$ (out of $286$) for which we have reliable market data for the whole period. Finally, we do not try to estimate possible transaction costs, but we compute the gross P\&L. In particular we compute the sign of the total P\&L at the above dates for the group of prelock and the group of newcomers.

The left panel of Figure \ref{fig:pl} shows, for the two groups, the fraction of inactive traders, computed as those with P\&L exactly equal to zero. We notice that a sizable fraction of the prelock investors were inactive after the lockdown, while a very small fraction of the newcomers were inactive: these are likely concentrating their trading in the $286-214=72$ very low cap stocks not considered in our analysis. This result indicates that a large fraction of the prelock investors essentially did not trade after the lockdown, being replaced by the newcomers, corroborating the results of the entry-exit dynamics discussed at the end of Section \ref{sec:entryexit}.

The right panel of Figure \ref{fig:pl} shows the fraction of investors with positive P\&L with respect to the number of active investors. We immediately notice that at all dates, the group of newcomers has a larger fraction of investors with positive profits than the group of prelock. This difference is maximal at the last date (September 29, 2021) when 85\% of the newcomers (versus 68\% of the active prelock investors) had a positive profit. This is a strong evidence that the newcomers are more or equally skilled than traditional (prelock) investors and cast some doubts on the hypothesis that gambling and sensation seeking were the main drivers of the arrival of new investors.

\section{Has the population of retail traders changed permanently?}\label{sec:shift}

As seen above, even restricting to the relatively short lockdown period, a large fraction of the newcomers continued to be active in trading Italian stocks also after the mobility restrictions have been progressively relaxed. This suggests that the extraordinary event of Covid has permanently modified the composition of the population of investors in Italian stocks. In particular, the mobility restrictions, the increased familiarity with IT platforms, and the availability of online trading environments have made more popular stock trading across households. 

\begin{figure}
\begin{center}	
	\includegraphics[width=0.45\textwidth]{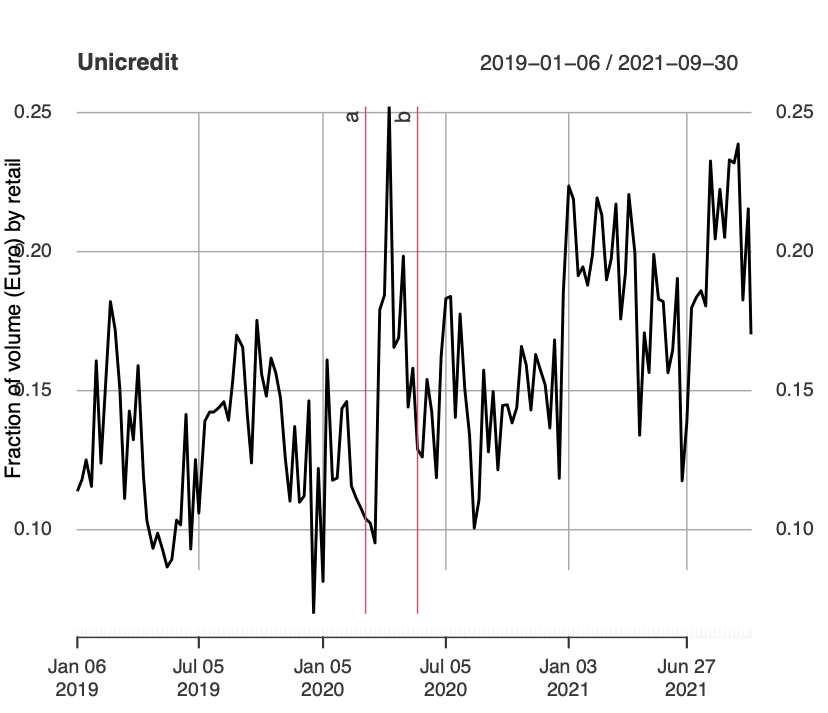}
	\includegraphics[width=0.45\textwidth]{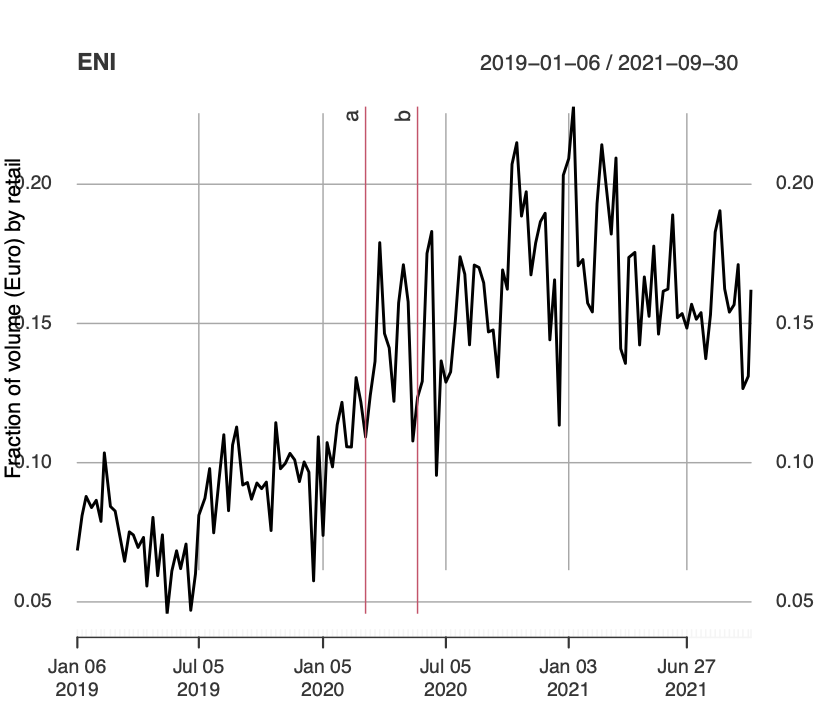}// 
	\includegraphics[width=0.45\textwidth]{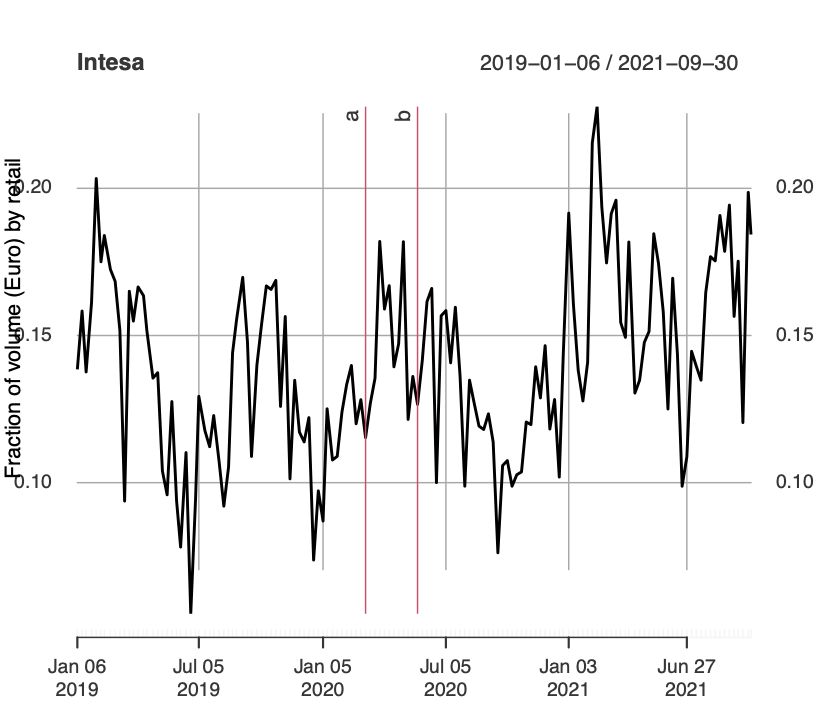}
	\includegraphics[width=0.45\textwidth]{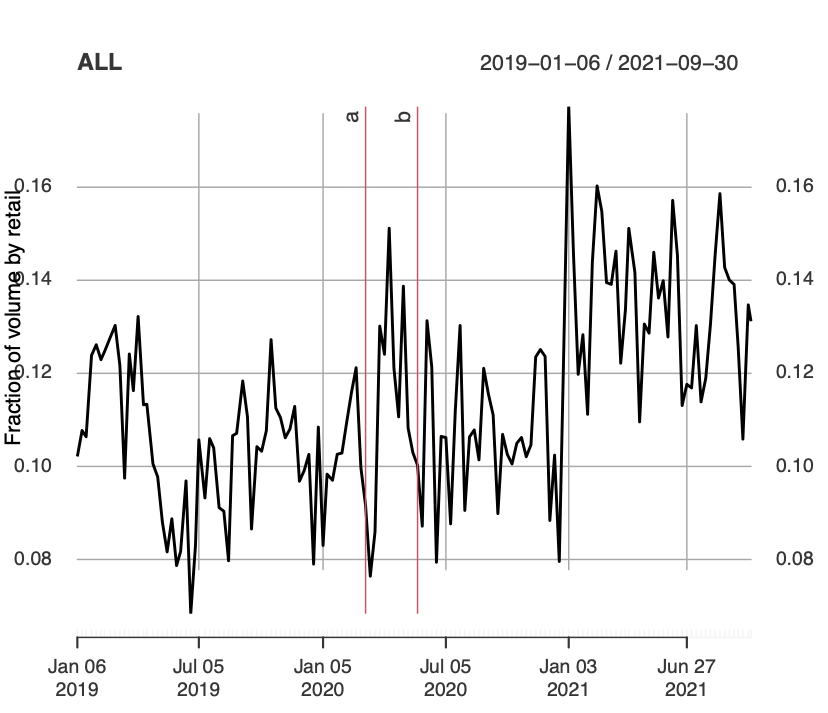}
\end{center}
\caption{Weekly fraction of volume (in euro) traded by retail investors as a total of the volume traded by retail investors and firms. The panels represent Unicredit (top left), ENI (top right), Intesa (bottom left), and all the stocks (bottom right).}\label{fig:frac}
\end{figure}

To test this idea, we measure for each week the fraction of trading volume due to households. Specifically, Figure \ref{fig:frac} shows the weekly fraction of volume (as usual measured in Euro) due to retail investors of the total volume traded by retail investors and firm. We notice that for some stocks (Unicredit and ENI) there has been a significant increase in the fraction of volume due to retail investors, i.e. their participation in the stock market has significantly increased during the pandemic and persists until the end of the investigated period. For few other stocks (Intesa) the effect seems less pronounced. Importantly, for the aggregate of all the stocks (bottom right panel) the increase is quite evident. This evidence suggests that (i) the pandemic has modified the structure of the composition of the investors population of Italian stocks with a stronger participation of retail investors; (ii) this shift is not uniform, thus some assets are likely experiencing a stronger shift.

\section{Conclusions and discussion}\label{sec:conclusion}
Covid pandemic has marked a shift in all the aspects of our lives and it is very likely that many of them will change permanently even after the end of the pandemic period. The study presented in this paper answers some questions about the impact of Covid on financial markets and, in particular, on the trading of Italian stocks.

Motivated by the great interest of Italian people towards stock markets at the beginning of the pandemic, we empirically investigate a unique dataset including the daily transactions of all investors operating in any of the Italian stocks, in the period from January 1, 2019 to September 30, 2021. Our findings suggest that the Italian stock market has undergone some changes from the lockdown and all the events triggered by the mobility restrictions have modified permanently its structure.

We show that the mobility restrictions have corresponded to an increase in the trading volume of all Italian stocks, likely due to the arrival of 185$k$ new retail investors, out of a total of about 563$k$ investors operating during the lockdown period. This has modified the ecology of the market in a number of aspects: (i) the new investors were on average younger than the typical investor present before the lockdown; (ii) they were more frequently male; (iii) they were (slightly) more skilled, as testified by the marked-to-market P\&L of their trading, which is positive more frequent than for other investors. Even if a small number of them has left the market by the end of the lockdown, we find that a large fraction has continued to operate, thus increasing permanently the population of retail traders. At the aggregate level, this is for example associated with a significant and permanent increase of the traded volumes by retail investors. Interestingly, likely due to the entrance of these new investors, the structure of the market has changed not only in numbers, but also in terms of trading behaviours. As in other markets, retail investors behave in aggregate as contrarian, buying when price goes down and vice versa. However, their net volume displayed a sharp transition when the first lockdown started and behaved as less contrarian during and after the lockdown, buying less than expected when stock prices were falling.

Several hypotheses can be put forward to explain such a change at the lockdown time and some of them have also been discussed in the recent literature. Among them one might consider:
\begin{itemize}
\item Investors entered because they observed a severe price drop and bet on a rebound of prices at the end of the acute phase of the pandemic, according to the so-called contrarian trading behavior \cite{Kaniel,pagano};
\item Investors interpreted buying and selling of stocks as gambling opportunity \cite{chiah2}; 

\item Investors had more time at their disposal due to the lockdown;
\item Investors had more money at disposal because of the reduced expenses for entertainment, gasoline, restaurant, etc.
\item Investors had more familiarity with electronic platforms, Fintech, etc., in part also because of remote working \cite{alber,fu,duchin};
\item Investors predicted a larger indebtedness of government due to extra health expenses and reduced taxes and thus predicted a decrease of government fixed income yield and increasing stock markets \cite{gormsen}; 
\end{itemize}

Our descriptive analysis does not allow to test rigorously which hypothesis should be rejected. We can however discuss how our empirical evidences are in support or against some of these hypotheses. 

The first hypothesis is essentially related to the contrarian behavior. We have seen that investors became less contrarian during the lockdown and most of them were still active many months after the end of the lockdown. Thus this opinion looks unlikely.

The persistence in the presence of most of the newcomers in financial markets after the end of the lockdown and their positive P\&L, also several months after the end of the lockdown, makes the gambling hypothesis less probable. On the contrary, the young age and the preponderance of males among newcomers might be in support of it, since the literature has shown that men have higher levels of engagement and problems in gambling than women \cite{gambling}. Interestingly, those who traded only during the lockdown were even younger and more frequently male, suggesting that for a fraction of investors in this restricted set gambling and sensation seeking could have been an important driver to enter the market. 

However, the demographic characteristics of the newcomers could also be used in support of other hypotheses: first, lockdown forced many women to conciliate work and family care at the same time and in the same place, leaving small possibility of being engaged in trading. Second, young people are naturally more familiar with electronic platforms and thus Covid lockdown might have represented an unique occasion to use these skills to enter the stock market, especially in times when the access to internet was essentially continuous, due also to smart working. Finally, the hypothesis on the differential trading in equities and government bonds could be tested with suitable data on the latter assets.

Even if, as said, we cannot rigorously rule out any of the above hypotheses, we believe that the persistent shift in the population of retail trader observed one year and half after the lockdown and the average profitability of the positions support the hypothesis of a regime change mostly driven by technological innovation. In this sense, Covid lockdown represented an accelerator of something that probably would have happened on a longer time scale (analogously to the adoption of smart working). Clearly, more analyses are needed to confirm or reject such hypothesis.

Covid pandemic had in general many negative consequences in our lives. It seems to have had a significant impact on financial markets. First, as highlighted in other works, it boosted  the digital transition in Finance, supporting the growth of FinTech and making the trading via online platforms of common use for investors. Here, for the first time, we show it had also the merit of attracting younger people that were not interested in investing before. And, even more importantly, a large part of them has continued to be present in the market. Since retail investors are fundamental to support liquidity provision, especially in turmoil periods, because of their contrarian behavior, and market liquidity is key for financial stability, it is possible to conclude that the renewal of the market population driven by the Covid pandemic is one signal of the health status of the Italian market. Nevertheless, the new and younger population shows a male prevalence, thus increasing the gender gap in financial activities. We will have to take on new actions and tailored policies in order to reduce the gender gap in finance, in particular by means of new investment in financial education for young people.

\begin{thebibliography}{100}



\bibitem{amf} AMF (2020). Report on retail investor behaviour during the Covid-19 crisis.
\href{https://www.amf-france.org/sites/default/files/2020-04/retail_investors_equities_march_2020_en.png}{https://www.amf-france.org/sites/default/files/2020-04/retail\_investors\_equities\_march\_2020\_en.png}

\bibitem{consobreport2021} CONSOB (2021). Report on financial investments of Italian households. \href{https://www.consob.it/web/consob-and-its-activities/report-on-investments-households}{https://www.consob.it/web/consob-and-its-activities/report-on-investments-households}

\bibitem{esma}ESMA (2022), Key Retail Risk Indicators for the EU single market, ESMA TRV Risk Analysis (forthcoming).

\bibitem{ft} Financial Times (2021). Rise of the retail army: the amateur traders transforming markets. \href{https://www.ft.com/content/7a91e3ea-b9ec-4611-9a03-a8dd3b8bddb5}{https://www.ft.com/content/7a91e3ea-b9ec-4611-9a03-a8dd3b8bddb5}

\bibitem{alber}Alber, N., and Dabour, M. (2020). The dynamic relationship between FinTech and social distancing under COVID-19 pandemic: Digital payments evidence. International Journal of Economics and Finance, 12(11).

\bibitem{baltakiene}  Baltakiene, M., Baltakys, K., Kanniainen, J., Pedreschi, D., and Lillo, F. (2019). Clusters of investors around initial public offering. Palgrave Communications, 5(1), 1-14.

\bibitem{odean} Barber, B. M., and Odean, T. (2000). Trading is hazardous to your wealth: The common stock investment performance of individual investors. Journal of Finance, 55(2), 773-806.

\bibitem{Barber} Brad M. Barber,  Yi-Tsung Lee,  Yu-Jane Liu,  Terrance Odean. Just How Much Do Individual Investors Lose by Trading? The Review of Financial Studies, 22, 609-632 (2009)

\bibitem{Odean2021} Barber, B. M., Lin, S., and Odean, T. (2021). Resolving a paradox: Retail trades positively predict returns but are not profitable. Available at SSRN 3783492.

\bibitem{Boehmer} Boehmer, E., Jones, C. M., Zhang, X., and Zhang, X. (2020). Tracking retail investor activity. Journal of Finance, Forthcoming.

\bibitem{tvreg} Casas, I., and Fernandez-Casal, R. (2019). tvReg: Time-varying coefficient linear regression for single and multi-equations in R. Available at SSRN 3363526.

\bibitem{chiah}Chiah, M., and Zhong, A. (2020). Trading from home: The impact of COVID-19 on trading volume around the world. Finance Research Letters, 37, 101784.

\bibitem{chiah2}Chiah, M., Tian, X., and Zhong, A. (2022). Lockdown and retail trading in the equity market. Journal of Behavioral and Experimental Finance, 33, 100598.

\bibitem{dorn} Dorn, D. and Sengmueller, P. (2009) Trading as entertainment? Management Science 55(4) 591-603.

\bibitem{duchin}Duchin, R., and Harford, J. (2021). The Covid-19 Crisis and the Allocation of Capital. Journal of Financial and Quantitative Analysis, 56(7), 2309-2319.

\bibitem{fu}Fu, J., and Mishra, M. (2020). The global impact of COVID-19 on FinTech adoption. Swiss Finance Institute Research Paper, (20-38).

\bibitem{gao} Gao, X., and Lin, T., (2015) Do individual investors treat trading as a fun and exciting gambling activity?
Evidence from repeated natural experiments The Review of Financial Studies, 28(7) 2128-2166.

\bibitem{gormsen}Gormsen, N. J., and Koijen, R. S. (2020). Coronavirus: Impact on stock prices and growth expectations. The Review of Asset Pricing Studies, 10(4), 574-597.

\bibitem{overconfidence} Grinblatt, M., and Keloharju, M. (2009). Sensation seeking, overconfidence, and trading activity. Journal of Finance, 64(2), 549-578.


\bibitem{hale} Hale, T., Angrist, N., Goldszmidt, R. et al. A global panel database of pandemic policies (Oxford COVID-19 Government Response Tracker). Nature Human behaviour 5, 529–538 (2021).

\bibitem{han} Han, H., Jian, W., Wang, N., and Zhao, B. (2014) Trading for Status, The Review of Financial Studies, 27(11) 3171-3212.

\bibitem{Kaniel} Kaniel, R., Saar, G., and Titman, S. (2008). Individual investor trading and stock returns. The Journal of finance, 63(1), 273-310.
 
\bibitem{news} Lillo, F., Miccichè, S., Tumminello, M., Piilo, J., and Mantegna, R. N. (2015). How news affects the trading behaviour of different categories of investors in a financial market. Quantitative Finance, 15(2), 213-229.
 
 \bibitem{ozik} Ozik, G., Sadka, R., and Shen, S. (2021). Flattening the illiquidity curve: Retail trading during the COVID-19 lockdown. Journal of Financial and Quantitative Analysis, 56(7), 2356-2388.
 
 \bibitem{pagano} Pagano, M. S., Sedunov, J., and Velthuis, R. (2021). How did retail investors respond to the COVID-19 pandemic? The effect of Robinhood brokerage customers on market quality. Finance Research Letters, 43, 101946.
 
  \bibitem{gambling} Stoltenberg, S. F., Batien, B. D., and Birgenheir, D. G. (2008). Does gender moderate associations among impulsivity and health-risk behaviors?. Addictive behaviors, 33(2), 252-265.
 
 \bibitem{svn} Tumminello, M., Lillo, F., Piilo, J., and Mantegna, R. N. (2012). Identification of clusters of investors from their real trading activity in a financial market. New Journal of Physics, 14(1), 013041.
 
\bibitem{overconfidence2} Talwar, S., Talwar, M., Tarjanne, V., and Dhir, A. (2021). Why retail investors traded equity during the pandemic? An application of artificial neural networks to examine behavioral biases. Psychology \& Marketing, 38(11), 2142-2163.


\end{thebibliography}
  \end{document}